\documentclass[a4paper,fleqn,usenatbib]{mnras}
\usepackage{ae,aecompl}

\usepackage{amsmath}
\usepackage{graphicx}
\usepackage{amssymb}
\usepackage{color}

\newcommand{\longcomet}{67P/Churyumov-Gerasimenko}
\newcommand{\shortcomet}{67P/C-G}
\newcommand{\HiiO}{\mathrm{H_2O}}
\newcommand{\COii}{\mathrm{CO_2}}
\newcommand{\CO}{\mathrm{CO}}
\newcommand{\Oii}{\mathrm{O_2}}

\newcommand{\voutflow}[1]{u_{#1,0}}
\newcommand{\mission}[3]{#1_{#2\mathrm{#3}}}
\newcommand{\helio}{{r_{\mathrm{h}}}}
\newcommand{\misbegin}{-372}
\newcommand{\misend}{390}
\newcommand{\datebegin}{August 6th 2014}
\newcommand{\dateend}{September 5th 2016}
\newcommand{\winbegin}{-330}
\newcommand{\winend}{-310}
\newcommand{\scraft}{spacecraft}

\newcommand{\nine}{nine}
\newcommand{\Nine}{Nine}

\title[Surface localization of gas sources] {Surface localization of
  gas sources on comet \longcomet{} based on DFMS/COPS data}
\author[Matthias L\"auter et al]{
  Matthias L\"auter,$^{1}$\thanks{E-mail: laeuter@zib.de} Tobias
  Kramer,$^{1,2}$ Martin Rubin$,^{3}$ Kathrin Altwegg$^{3}$ \\
  $^{1}$Zuse Institute Berlin, 14195 Berlin, Germany\\
  $^{2}$Department of Physics, Harvard University, Cambridge, MA, USA\\
  $^{3}$University of Bern, Physikalisches Institut, 3012 Bern, Switzerland}

\date{Accepted XXX. Received YYY; in original form ZZZ}

\pubyear{2018}

\begin{document}
\label{firstpage}
\pagerange{\pageref{firstpage}--\pageref{lastpage}}
\maketitle

\begin{abstract}

We reconstruct the temporal evolution of the source distribution for
the four major gas species $\HiiO$, $\COii$, $\CO$, and $\Oii$ on the
surface of comet \longcomet{} during its 2015 apparition.
The analysis applies an inverse coma model and fits to data between
\datebegin{} and \dateend{} measured with the Double Focusing Mass
Spectrometer (DFMS) of the Rosetta Orbiter Spectrometer for Ion and
Neutral Analysis (ROSINA) and the COmet Pressure Sensor (COPS).
The spatial distribution of gas sources with their temporal variation
allows one to construct surface maps for gas emissions and
to evaluate integrated productions rates.
For all species peak production rates and integrated productions rates
per orbit are evaluated separately for the northern and the
southern hemisphere.
The \nine{} most active emitting areas on the comet's surface are
defined and their correlation to emissions for each of the species is
discussed.

\end{abstract}

\begin{keywords}
comets: individual: \longcomet{} -- methods: data analysis
\end{keywords}

\section{Introduction}

Solar radiation triggers the activity of comets as they approach the
inner solar system and start to release a mixture of different
volatiles and solid dust grains.
The Rosetta mission has studied the nucleus and the
environment of comet \longcomet{} (\shortcomet{}).
The suite of instruments examining volatiles and dust on-board the
\scraft{} incorporates ROSINA, VIRTIS, MIRO, GIADA, COSIMA and OSIRIS
(\cite{Schulz2009}).
Optical instruments probe the integrated intensity of dust and gas
along the line of sight, while the mass spectrometers and pressure
sensors measure the local composition and density in the coma at the
momentary \scraft{} position.
All measurement data must be embedded in a global coma model for
interpretation and reconstruction of the three-dimensional volume
density.

Analytical coma models starting with \cite{Haser1957} are complemented by
computational models reflecting the flow dynamics, illumination
conditions, and the complex non-spherical shape of the nucleus on various levels of complexity.
The reproduction of measurements necessitates the determination of
unknown surface parameters from observations.
\cite{Marshall2017} incorporate MIRO data into a local effective Haser
model based on projections into nadir direction to attribute
production rates to separated surface regions in their Fig.~6.
Based on three-dimensional shape models,
\cite{Bieler2015}, \cite{Marschall2015}, and \cite{Marschall2017}
introduce gaskinetic models
(Direct Simulation Monte Carlo codes).
\cite{Bieler2015} apply a parameter fit for a latitudinal dependence
of the gas activity.
\cite{Fougere2016}, \cite{Fougere2016a}, and later \cite{Hansen2016}
apply an inverse approach to an analytical gas model
(\cite{Fougere2016}, Eq.~(3)) and assimilate DFMS data to 25
coefficients of spherical harmonics.
These local inhomogeneities define the inner boundary condition of
their DSMC model.
\cite{Kramer2017} introduce a different simplified gas model and fit surface
production rates on $10^4$ surface elements to COPS density
data.

Here, we analyze the species resolved coma of \shortcomet{} and trace
the evolution of $\sim$~4000 gas emitters on the nucleus
every 14~days for more than $\pm 350$~days around perihelion.
This corresponds to heliocentric distances in the range of
$3.5-1.24$~au.
Our model connects individual gas-density observations with
limited spatial/temporal resolution to the surface activity across the
entire nucleus.
The input data to the model is the combined ROSINA COPS and DFMS data set.
The data processing is detailed in Sect.~\ref{sec:data}.
By parameterizing the measured density in terms of surface
emitters following \cite{Kramer2017}, we reconstruct the temporal
evolution of the gas emission rates of the four major volatiles
$\HiiO$, $\COii$, $\CO$, and $\Oii$ (Sect.~\ref{sec:model}).
In addition, our method determines the spatial distribution of the
species on the surface and reveals different production rates and ice
distributions on the northern and southern hemispheres
(Sect.~\ref{sec:qtotal}).
The production rates are compared to the MIRO data presented by
\cite{Marshall2017}, to the RTOF data by \cite{Hoang2017}, and with
the COPS analysis by \cite{Hansen2016}.
The localization of the most active emitting areas in
Sect.~\ref{sec:surface} is in good agreement to \cite{Hoang2017} and \cite{Kramer2017}.
This activity pattern shows a high correlation ($0.7$) to active gas
emitters with short living dust locations derived from OSIRIS and
NAVCAM images by \cite{Vincent2016}.
We recover ice-rich spots for $\HiiO{}$ and $\COii{}$ found
by \cite{Filacchione2016} and \cite{Fornasier2016}.
Sect.~\ref{sec:summary} provides a summary of our findings
and describes possible contributions to first-principle modeling
of cometary activity.

\section{Processing and interpolation of DFMS data}
\label{sec:data}

The Rosetta Orbiter Spectrometer for Ion and Neutral Analysis (ROSINA)
consisted of the two mass spectrometers DFMS (Double Focusing Mass
Spectrometer) and RTOF (Reflectron-type Time Of Flight) and COPS, the
COmet Pressure Sensor, see \cite{Balsiger2007}.
COPS measured the total gas density at the location of the
Rosetta spacecraft whereas the two mass spectrometers obtain the
relative abundances of the volatiles including the major parent
species $\HiiO$, $\COii$, $\CO$, and $\Oii$.
Combining COPS with the DFMS mass spectrometer, total abundances at
Rosetta can be derived (for details see \cite{Gasc2017}).
Our measured data considers the latest detector aging model as
described by \cite{Schroeder2018}.

Rosetta moved rather slowly with respect to the comet (typically
$<1$~m/s).
However, the comet rotates once per $\sim 12$~hours and the
combination of the comet's shape and tilt in the rotation axis led to
a complex variation of the measured abundances, both in relative and
absolute numbers (see \cite{Fougere2016}).

\begin{figure*}
\includegraphics[width=0.95\textwidth,draft=false]{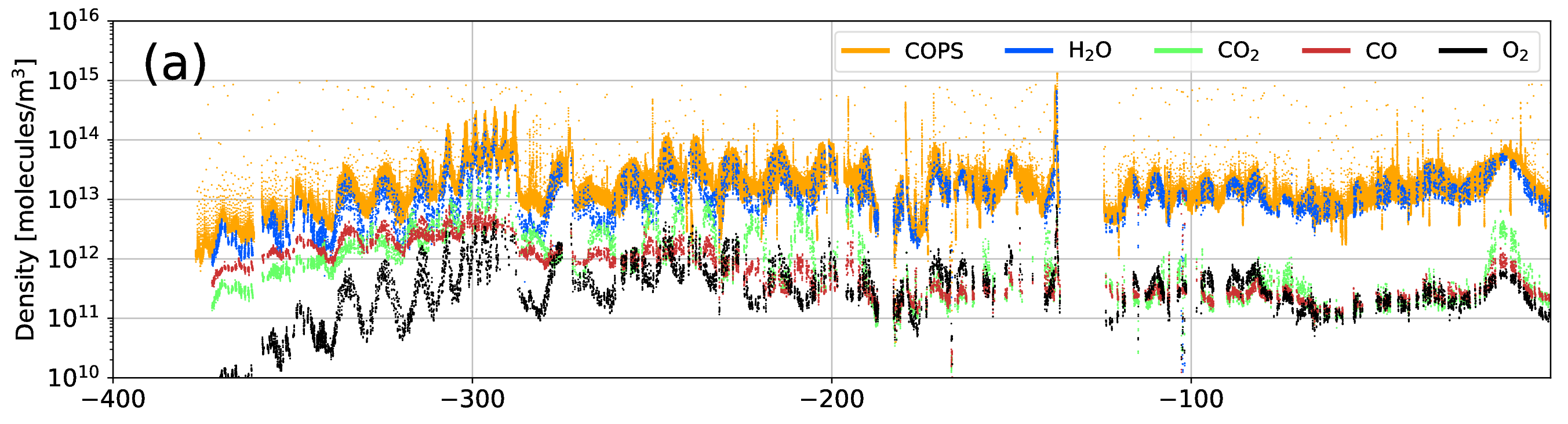}\\
\includegraphics[width=0.95\textwidth,draft=false]{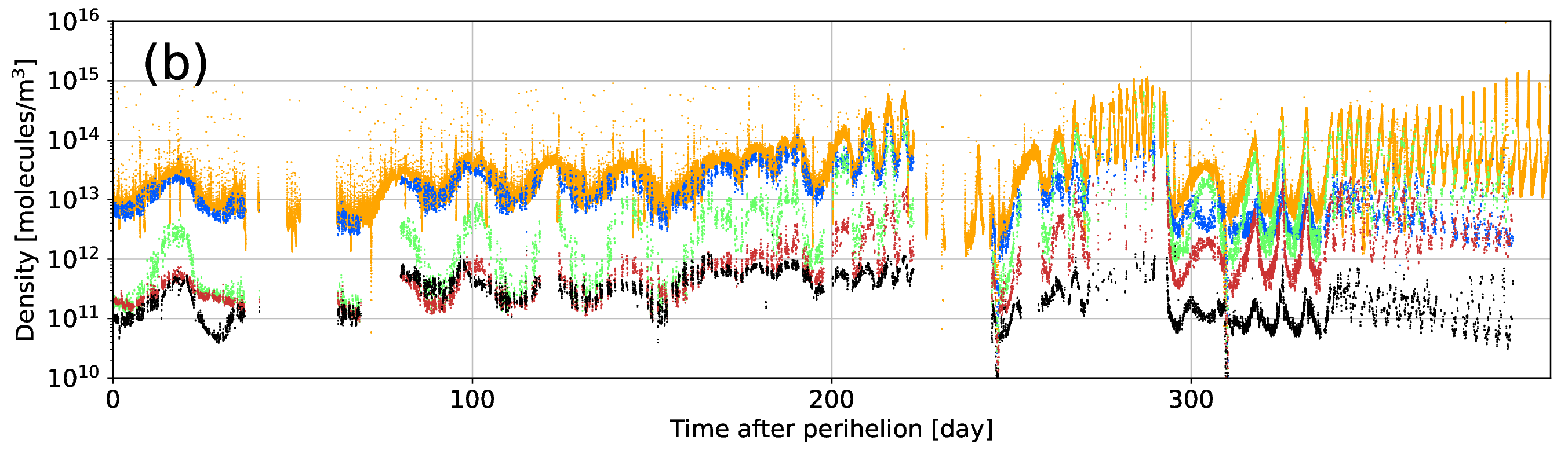}
\caption{Observations at the positions of Rosetta,
 COPS data at times $T_{\mathrm{COPS}}$, DFMS data at times $T_{\mathrm{DFMS}}$,
 (a) time interval $(-400,0)$~days, (b) time interval $(0,400)$~days.}
\label{fig:copsdfms_all}
\end{figure*}

\begin{figure*}
\includegraphics[width=0.95\textwidth,draft=false]{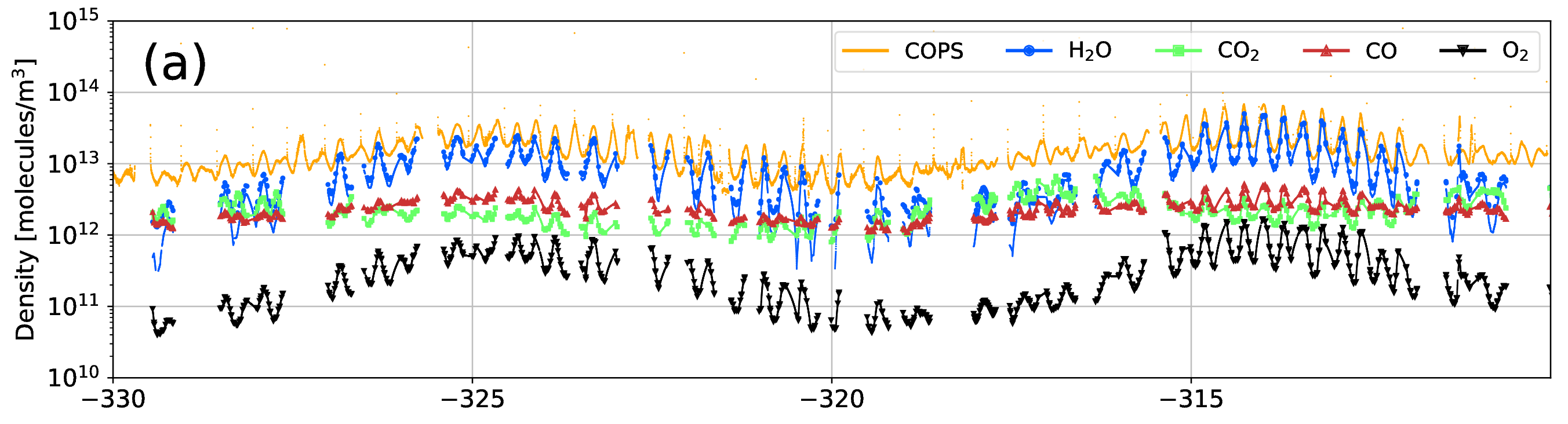}\\
\includegraphics[width=0.95\textwidth,draft=false]{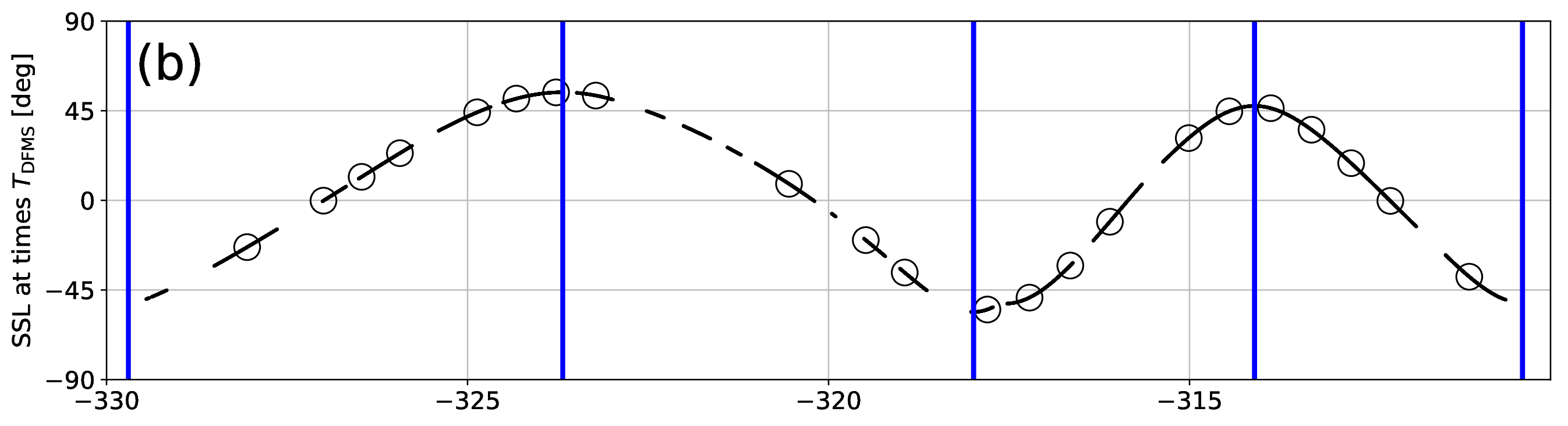}\\
\includegraphics[width=0.95\textwidth,draft=false]{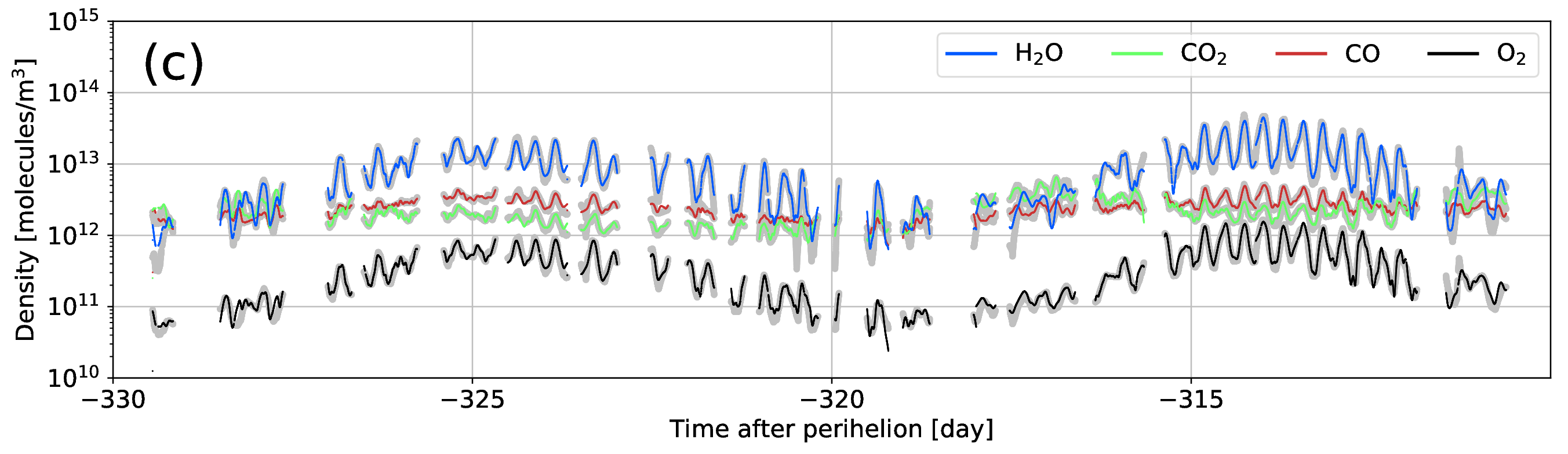}
\caption{(a) Observations at the positions of Rosetta,
  COPS data at times $T_{\mathrm{COPS}}$,
  DFMS data at times $T_{\mathrm{DFMS}}$ (filled symbols), and DFMS data in
  Eq.~\eqref{eqn:dfmsdata} interpolated at the times $T_{4\mathrm{h}}$
  (lines) in the time interval $(\winbegin,\winend)$~days. (b)
  sub-\scraft{} latitude (SSL) at the times $T_{4\mathrm{h}}$ (lines),
  $0^\circ$-meridian crossings of the \scraft{} (circles) and
  end points of time intervals $I_j$ (vertical lines).  (c)
  Inverse model fits at times $T_{4\mathrm{h}}$,
  The DFMS data from (a) is plotted in gray lines, the dominant species
  are from top to bottom $\HiiO$, $\CO$, $\COii$, and $\Oii$ at day $-325$.}
\label{fig:copsdfms_detail}
\end{figure*}

The total gas density at Rosetta's location is monitored by the COPS
instrument throughout most of the mission with a time resolution of
one minute.
The times of measurements are denoted by $T_{\mathrm{COPS}}$.
Our dataset includes 949381 COPS measurements and is depicted in
Fig.~\ref{fig:copsdfms_all}.
The measurements are taken between \datebegin{} and \dateend{},  
$(\misbegin{},\misend{})$~days from perihelion on August
13th 2015.
Negative values denote times before perihelion.
In addition to COPS, the DFMS instrument determines the relative
abundances of $\HiiO{}$, $\COii{}$, $\CO{}$, and $\Oii{}$ at a lower
time-resolution ($T_{\mathrm{DFMS}}$ denotes all times of measurements).
The DFMS dataset contains 32700 points, see
Fig.~\ref{fig:copsdfms_all}.
Fig.~\ref{fig:copsdfms_detail}(a) shows both data sets
in the exemplary time interval
$(\winbegin,\winend)$~days.
To increase the number of data points entering our DFMS coma model, 
we linearly interpolate the species resolved DFMS densities to the 
COPS times $T_{\mathrm{COPS}}$.
Spurious extrapolation artifacts are avoided by restricting the
interpolation to a $4$~h sized window around each point in
$T_{\mathrm{DFMS}}$, namely $T_{4\mathrm{h}} = \{ t \in
T_{\mathrm{COPS}}\,|\, t \in (t_{l},t_{r}), |t_{r}-t_{l}| <
4\mathrm{h}, t_{l},t_{r} \in T_{\mathrm{DFMS}} \}$.
The resulting 489009 interpolated densities are denoted by
\begin{equation}\label{eqn:dfmsdata}
\rho_{\HiiO{}}(t), \rho_{\COii{}}(t), \rho_{\CO{}}(t), \rho_{\Oii{}}(t),
\quad\text{for}\quad
t \in T_{4\mathrm{h}}.
\end{equation}
The different densities at the times $T_{\mathrm{COPS}}$,
$T_{\mathrm{DFMS}}$, and $T_{4\mathrm{h}}$ are depicted in
Fig.~\ref{fig:copsdfms_detail}(a).

\section{Reconstruction of the coma from local measurements}
\label{sec:model}

\begin{figure}
\includegraphics[width=0.475\textwidth,draft=false]{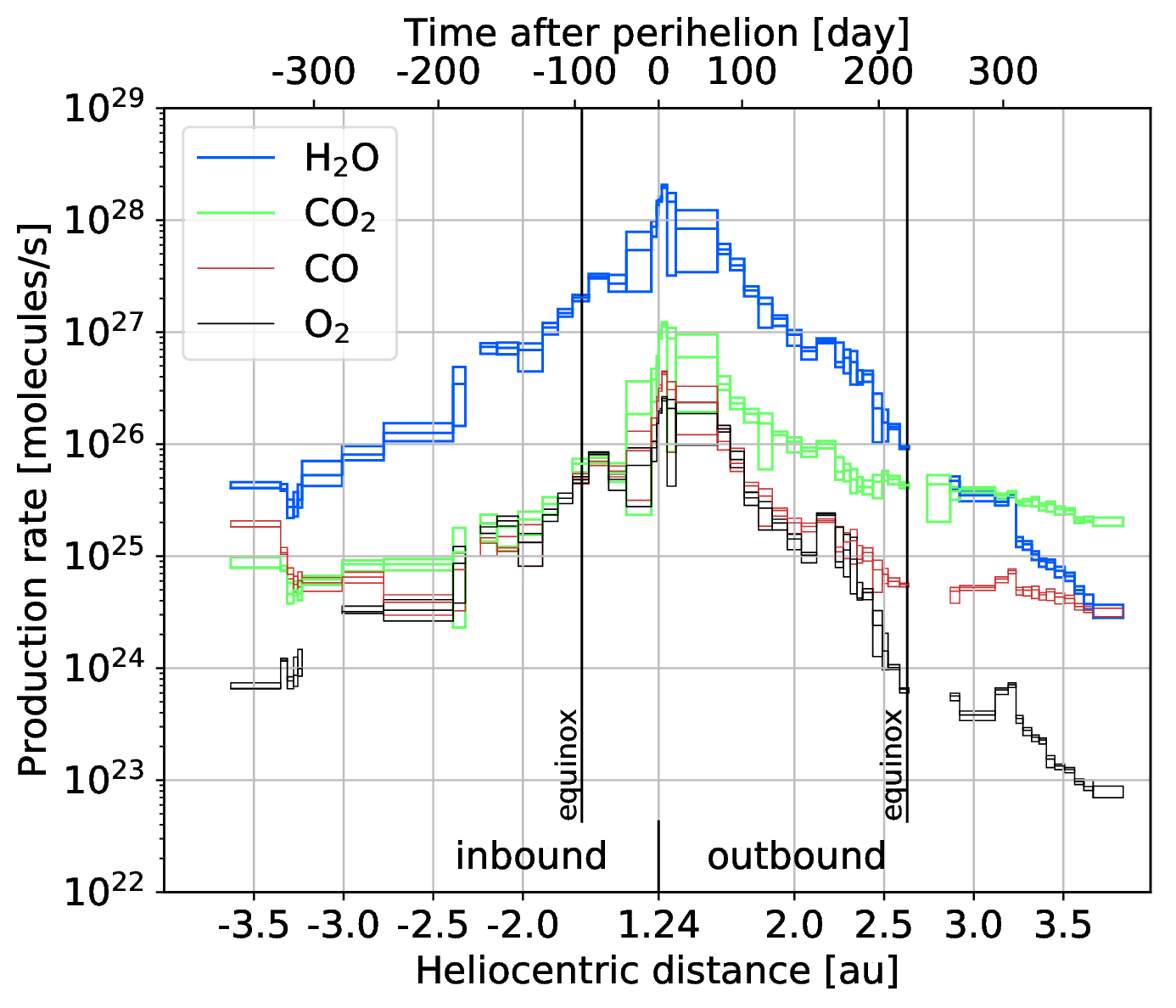}
\caption{Production rates $Q_s(t)$ for the species $s=\HiiO$, $\COii$,
  $\CO$, and $\Oii$ over time and heliocentric distance. The boxes denote the
  minimum, linear and maximum estimates due to varying \scraft{}
  surface coverage, see section~\ref{sec:model}.
  From top to bottom the dominant species are
  $\HiiO$, $\COii$, $\CO$, and $\Oii$ at outbound equinox.
}
\label{fig:qtimespecies}
\end{figure}

\begin{figure*}%
\includegraphics[width=0.5\textwidth,draft=false]{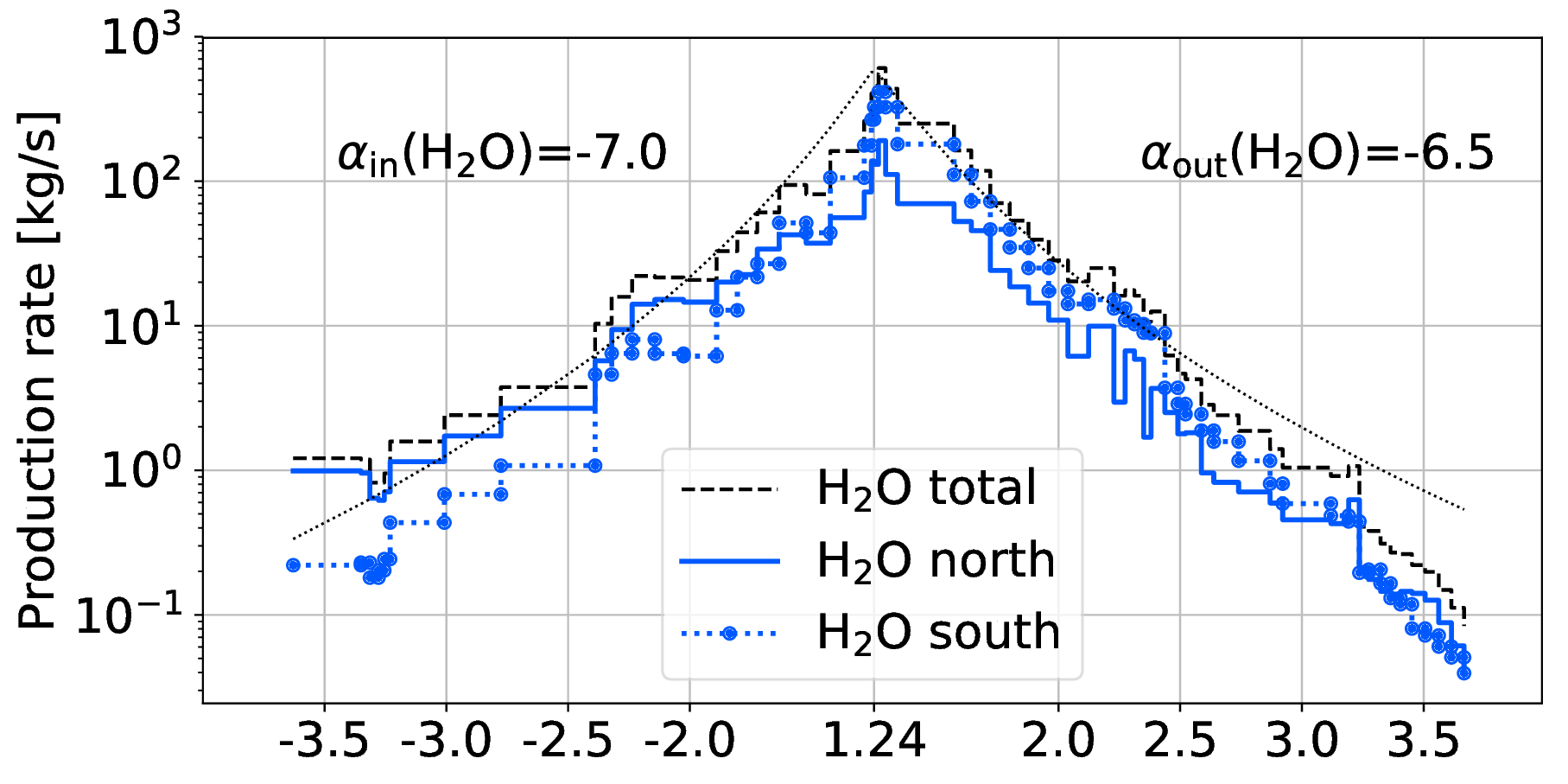}%
\includegraphics[width=0.5\textwidth,draft=false]{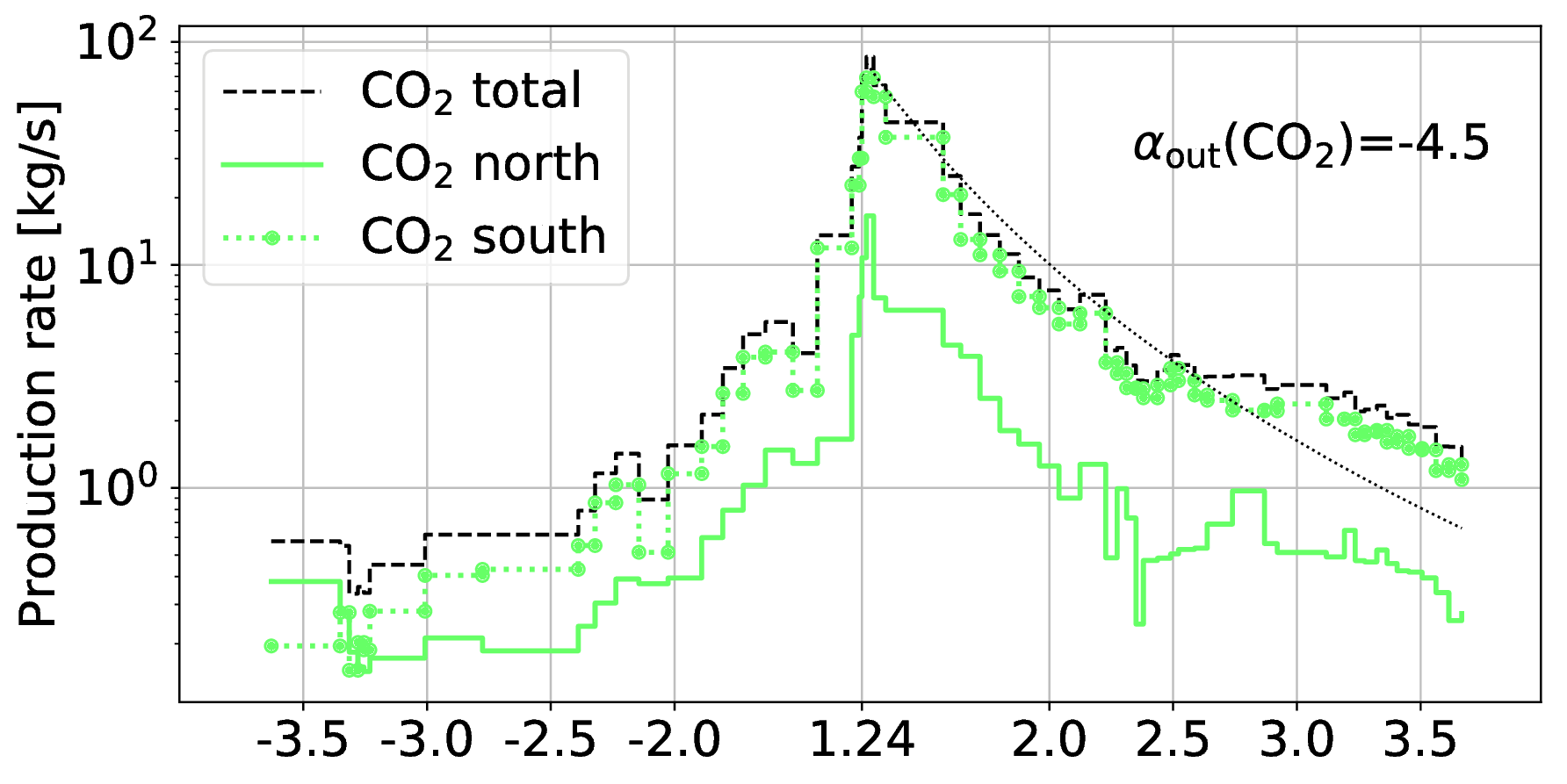}\\
\includegraphics[width=0.5\textwidth,draft=false]{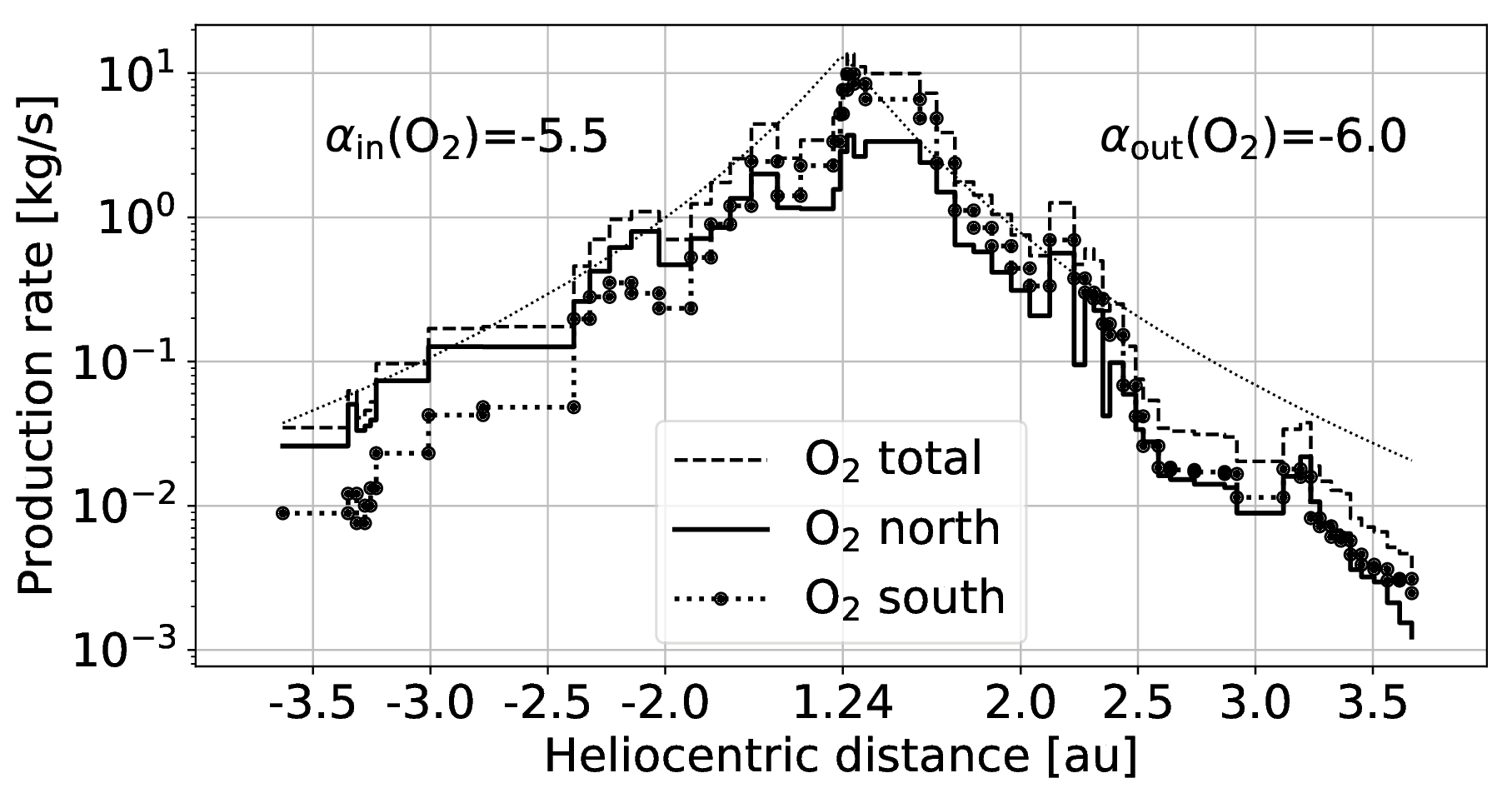}%
\includegraphics[width=0.5\textwidth,draft=false]{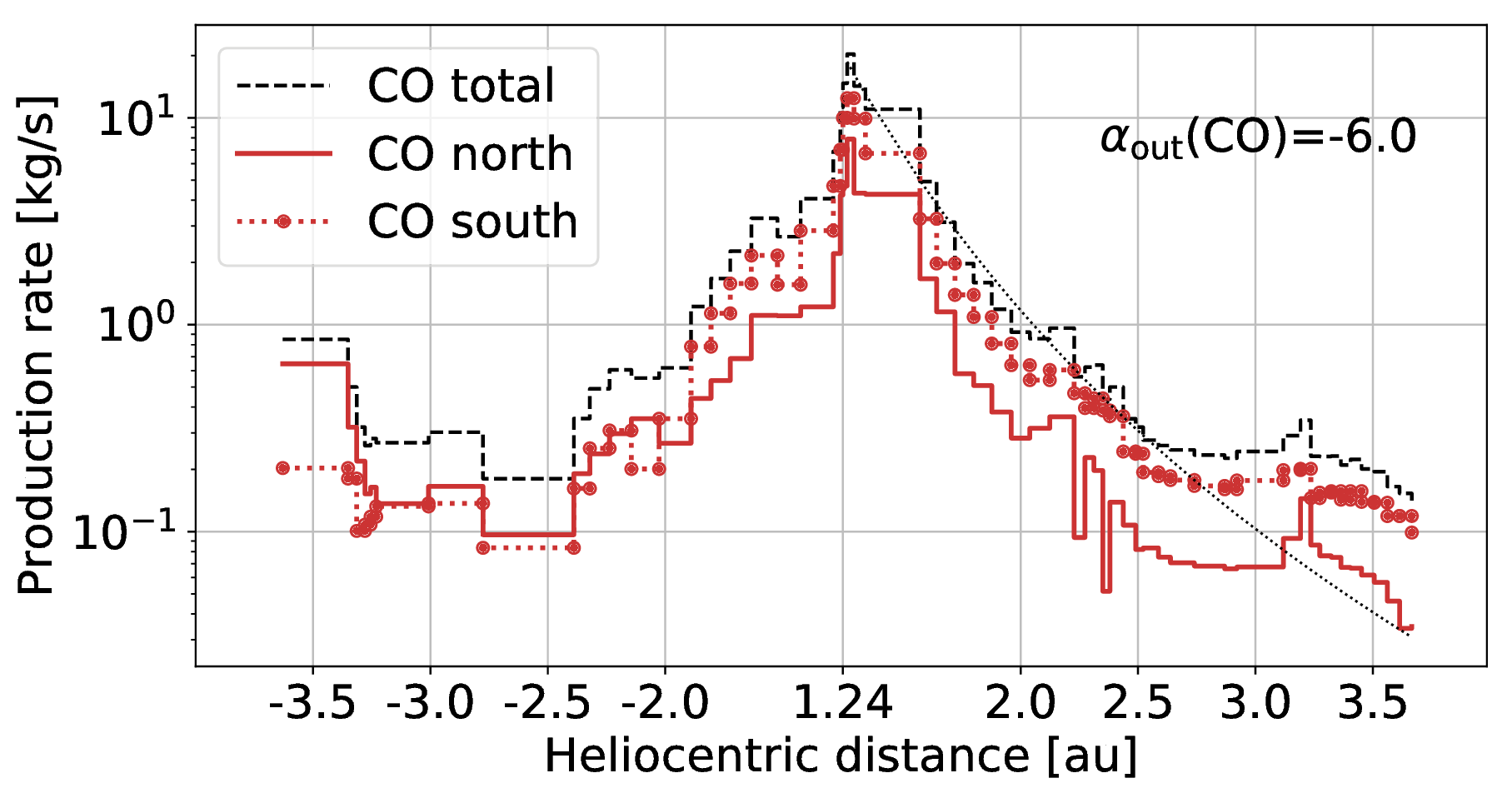}%
\caption{Production rates $\mission{Q}{s}{}(\helio)$ (split into
  northern, southern hemisphere and total) for the species $s=\HiiO$,
  $\COii$, $\CO$, and $\Oii$ as a function of heliocentric distance
  $\helio$, power-law fits $\mission{Q}{s}{}(\helio)\sim
  \helio^{\alpha(s)}$.}
\label{fig:qtimelinear}
\end{figure*}

The global reconstruction of the entire three-dimensional coma around
\shortcomet{} proceeds as a two-step process from the time-series of COPS
and DFMS measurements along the trajectory of Rosetta
and is based on the assignment of surface emission rates as described by
\cite{Kramer2017}.
First we run a forward model on a surface shape to build a global coma
model by assuming
equally strong emitting gas sources on each of the surface elements.
In the second step we apply the inverse model and adjust the emission
rates of each source to obtain the best match with the actually
measured DFMS/COPS data.
Systematic model uncertainties (insufficient observational sampling in
space or/and time) are discussed below.

The whole surface of the nucleus is approximated by a triangular mesh
with $N_\mathrm{E} = 3996$ equidistantly spaced surface elements,
leading to a spatial resolution of $110$~m on average.
The original shape model \cite{spcesamtp019} is remeshed using the
ACDVQ tracing tool by \cite{Valette2008a} and smoothed.
We have validated the method by performing the model inversion for
more and less detailed shape models.
The surface reconstructions from higher-resolution models are slightly
more scattered (see \cite{Kramer2017} for COPS data), but do not
change the regional results discussed here.

To follow the evolution of the emission rates as the comet orbits the
sun, we divide the complete time interval
$(\misbegin{},\misend{})$~days into $N_\mathrm{I}=58$ subintervals.
\begin{gather*}
\misbegin{} = t_0, t_1, ..., t_{N_\mathrm{I}} = \misend{}, \\
I_j = (t_{j-1},t_j),
\quad\text{for}\quad
j = 1, ..., N_\mathrm{I}
\end{gather*}
Each subinterval $I_j$ includes 8600 values from $T_{4\mathrm{h}}$ on
average and comprises typically $14$~days.
As an example, Fig.~\ref{fig:copsdfms_detail}(b) shows four
subintervals, each enclosing extremal sub-\scraft{} latitudes and five
or more comet rotations.
Because the data points need to constrain the parameters,
the complete determination of the $N_\mathrm{E}$ model parameters
(here: the surface emission rate) requires to have more data points
available (here: DFMS/COPS measurements).
The intervals $I_j$ are chosen such that the \scraft{} positions in
$I_j$ result in an almost complete coverage of the nucleus surface.
Surface sources with no flyover within the interval $I_j$ are set to zero
emission for the lower-bound estimate of the activity.

For building the forward model, we consider the approach of
\cite{Kramer2017} and introduce a model for a collisionless gas regime
in the coma.
Around perihelion and close to the nucleus, estimated gas densities of
up to $10^{18}~\mathrm{molecules}/\mathrm{m}^3$ result in mean free
paths of about $3$~m.
This value is considerably larger than the mean free paths considered
by \cite{Gombosi1986} (0.1-1)~m, \cite{Crifo2004} ($<1$~m), and
\cite{Tenishev2008} ($<1$~m) and results in higher Knudsen numbers
$>0.003$.
Away from perihelion and further away from the nucleus, the fast $\sim
1/r^2$ drop in gas density quickly leads to intermediate and
collisionless flow regimes.
From Fig.~2 in \cite{Finklenburg:2011}, we estimate the uncertainties
due to collisions at observational \scraft{} distances to be less than
25\% around perihelion, resulting in smaller contributions to the
model uncertainties compared to coverage and fitting errors.

On every surface element the model assumes a point source, which emits
gas with a displaced Maxwellian velocity distribution shifted by a
given mean velocity.
This leads to the analytical expression Eq.~(1) in \cite{Kramer2017}
for the density derived by \cite{Narasimha1962}.
The lateral expansion of the gas column perpendicular to the surface
normal is taken into account.
The modeled gas density at every space point around the nucleus arises
from a superposition of all surface emitters.
The accurate incorporation of the nucleus shape and the possibility to
assign multiple surface locations to a single gas measurement set our
model apart from a simple nadir mapping of data points.
The nadir method projects each \scraft{} measurement onto a single
point on the surface of the nucleus.

\begin{figure*}
\begin{tabular}{lcccc}
& view from north & view from south & view
  from (${0,-1,0}$) & view from (${0,1,0}$) \\
A&
\parbox{0.2\textwidth}{%
\includegraphics[width=0.2\textwidth,draft=false]{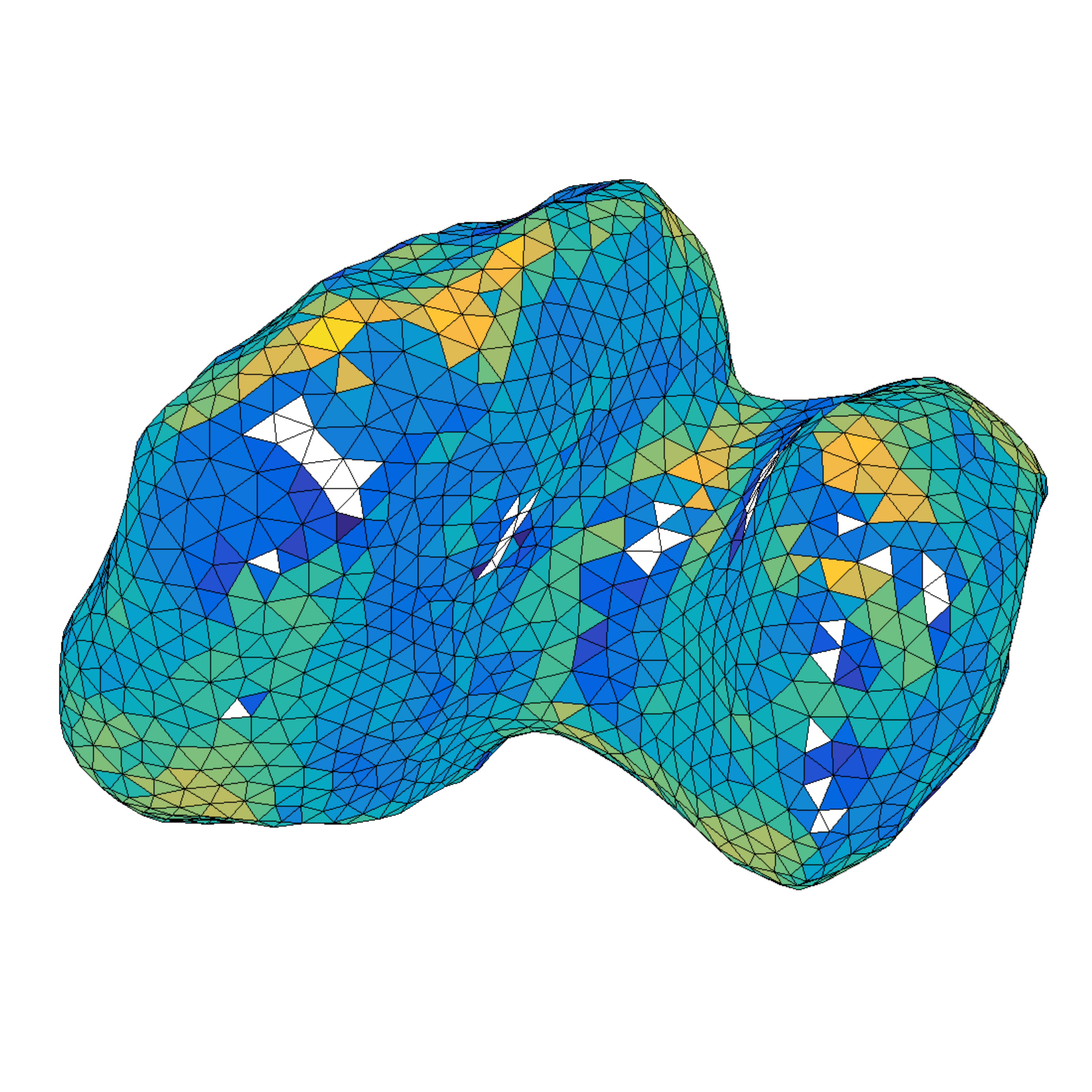}}&
\parbox{0.2\textwidth}{%
\includegraphics[width=0.2\textwidth,draft=false]{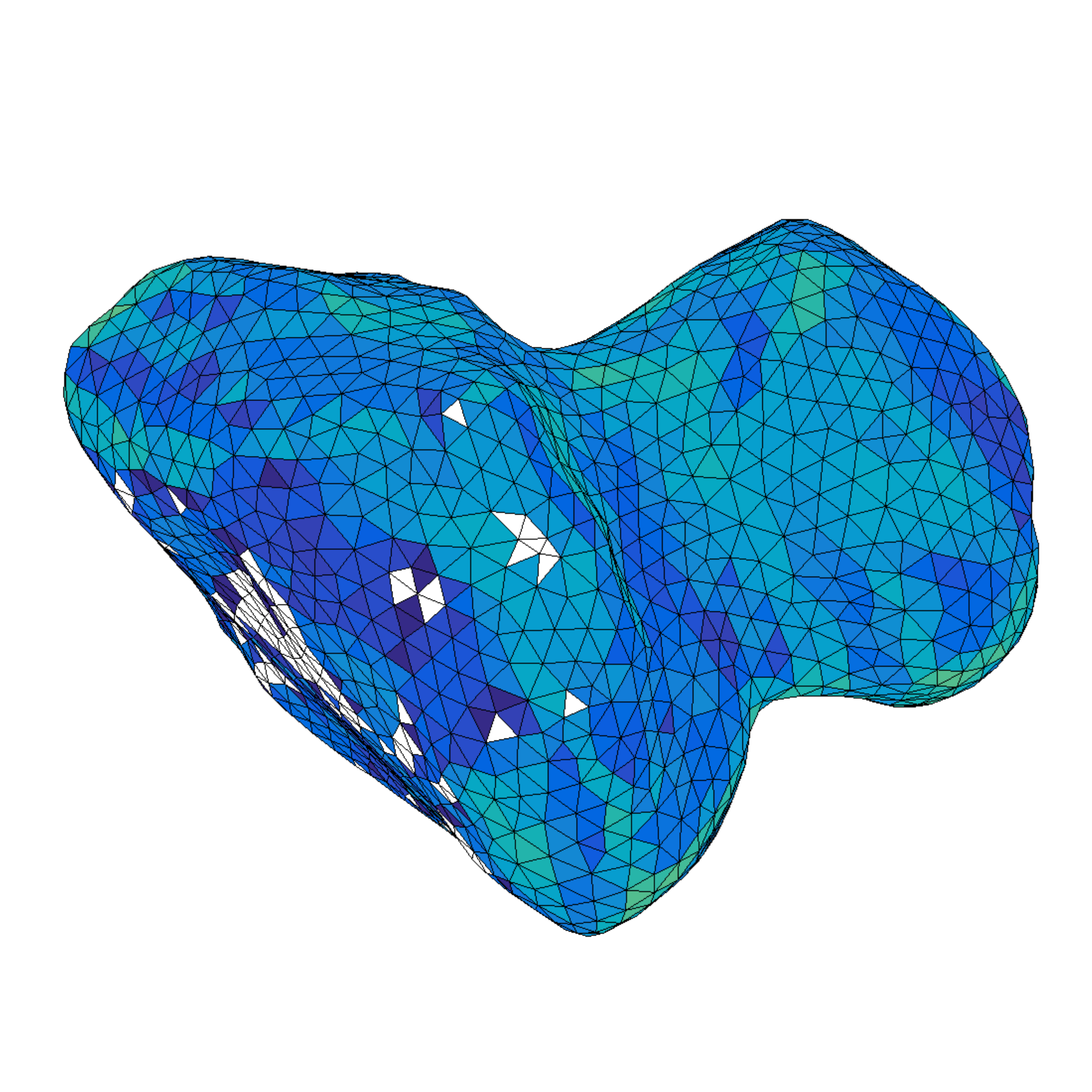}}&
\parbox{0.2\textwidth}{%
\includegraphics[width=0.2\textwidth,draft=false]{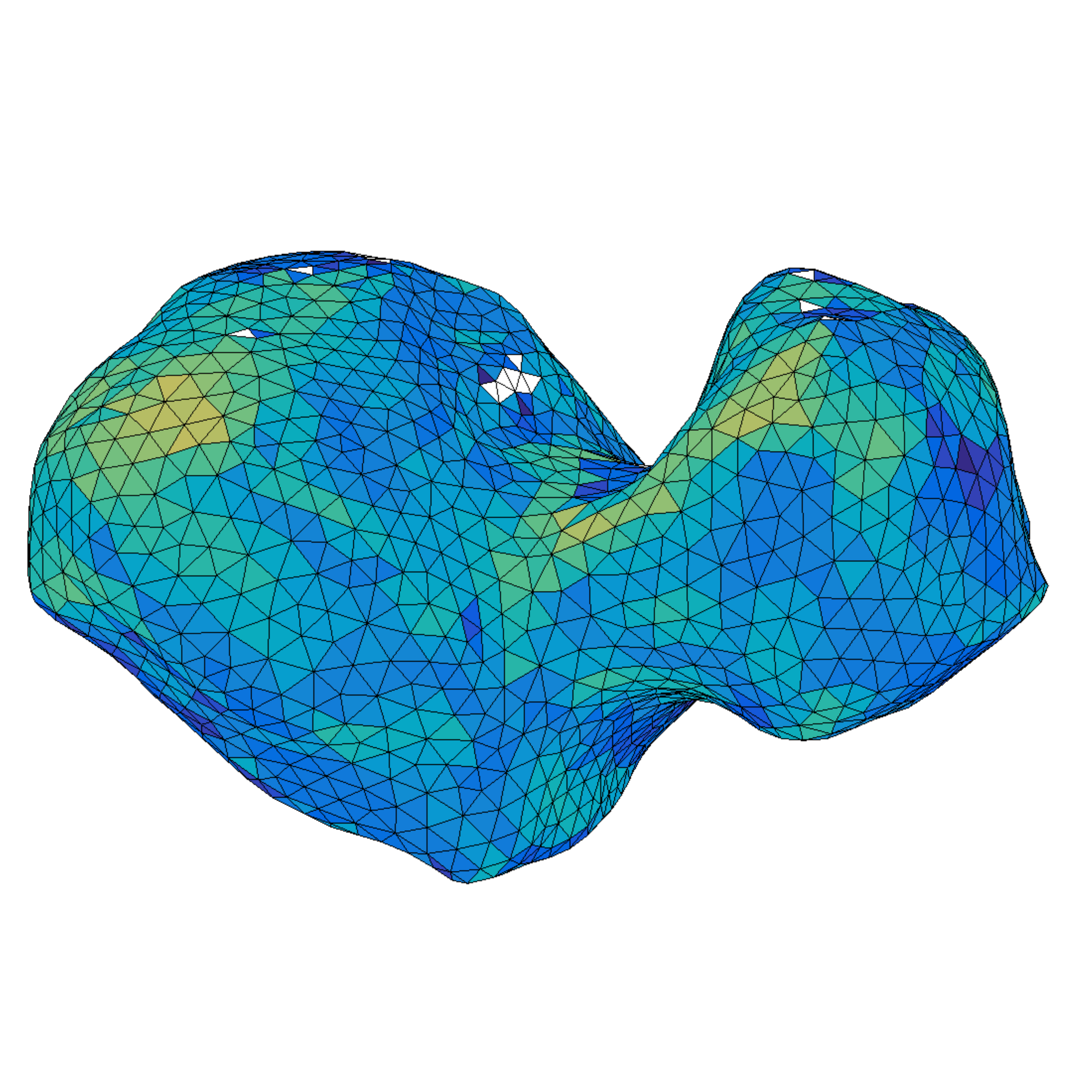}}&
\parbox{0.2\textwidth}{%
\includegraphics[width=0.2\textwidth,draft=false]{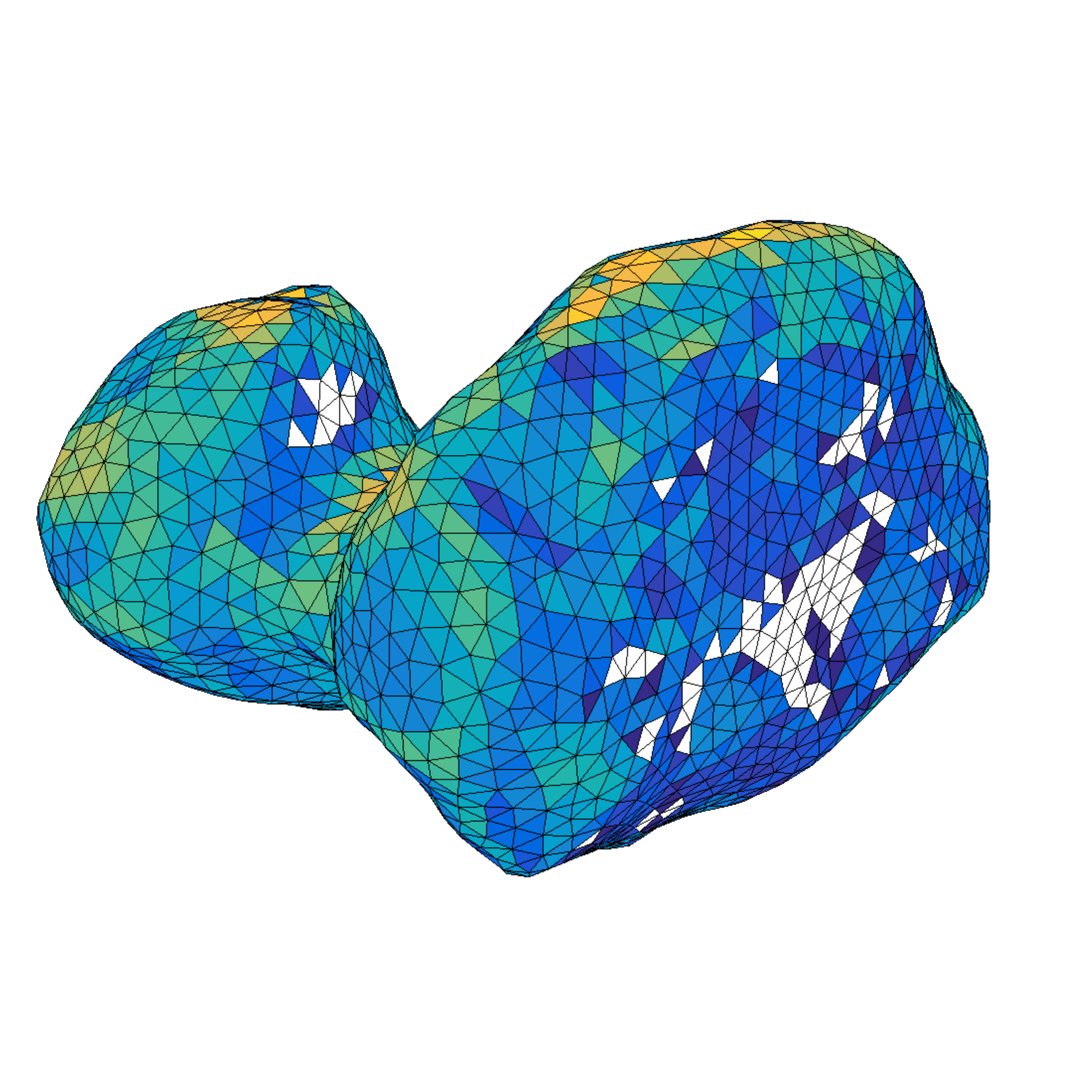}}\\
B&
\parbox{0.2\textwidth}{%
\includegraphics[width=0.2\textwidth,draft=false]{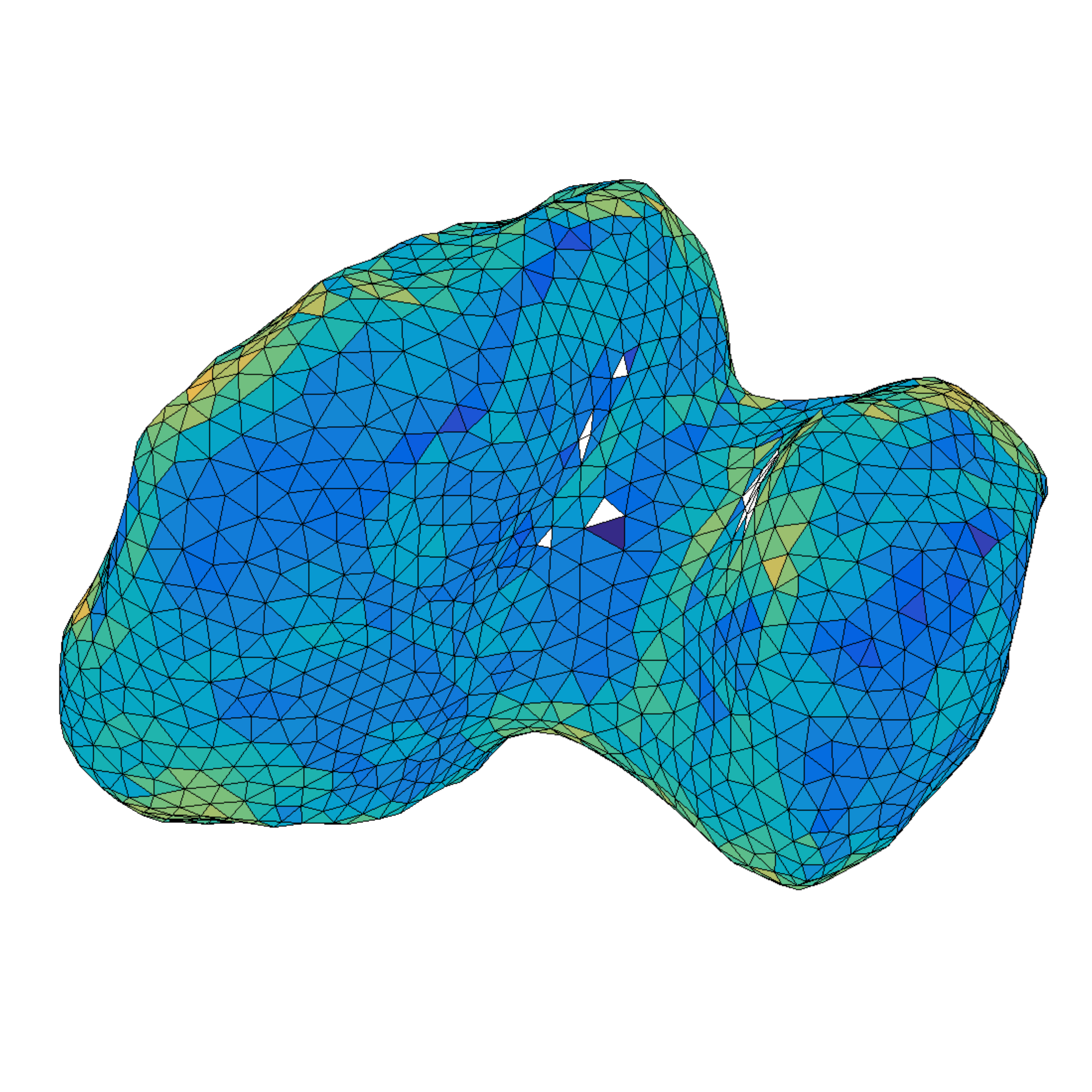}}&
\parbox{0.2\textwidth}{%
\includegraphics[width=0.2\textwidth,draft=false]{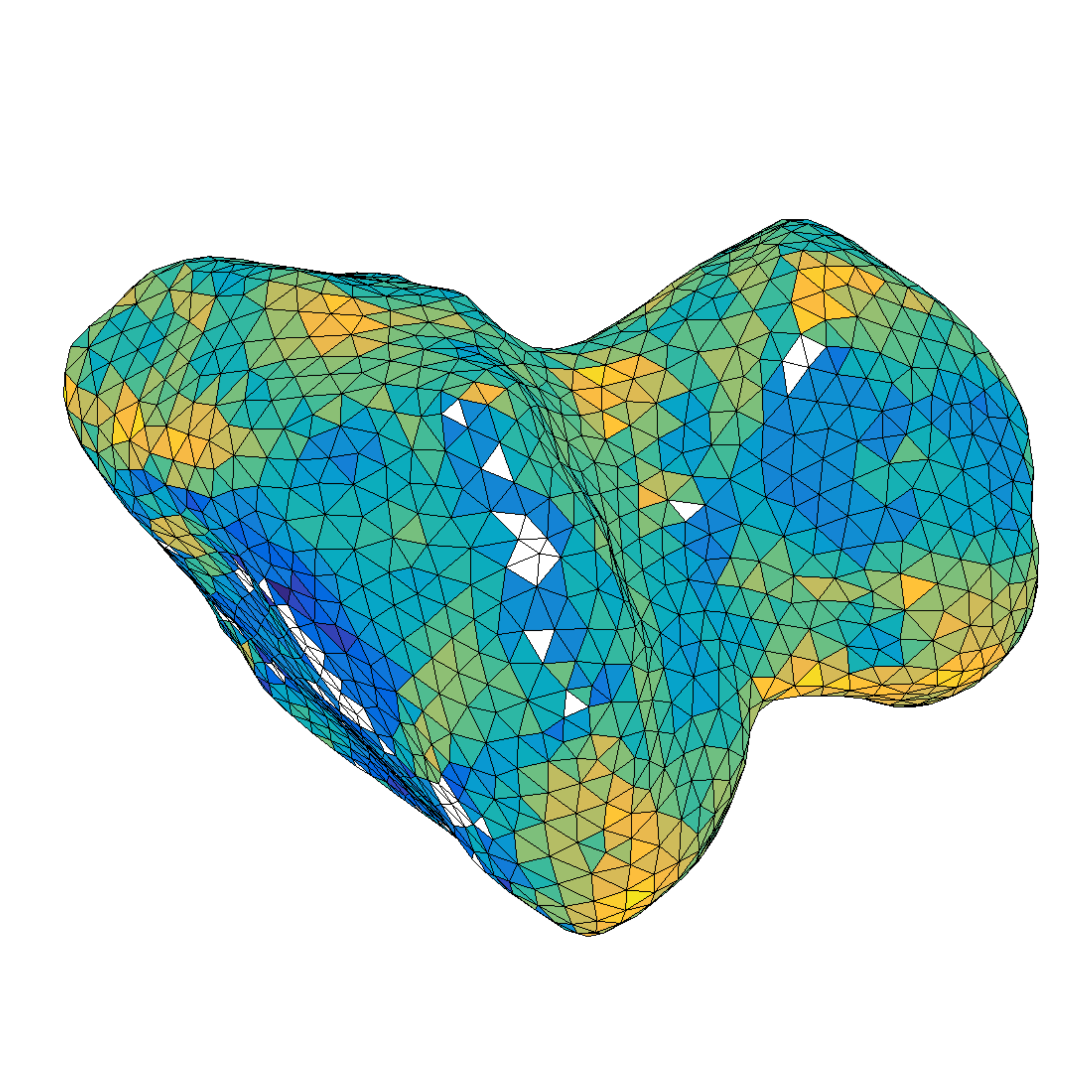}}&
\parbox{0.2\textwidth}{%
\includegraphics[width=0.2\textwidth,draft=false]{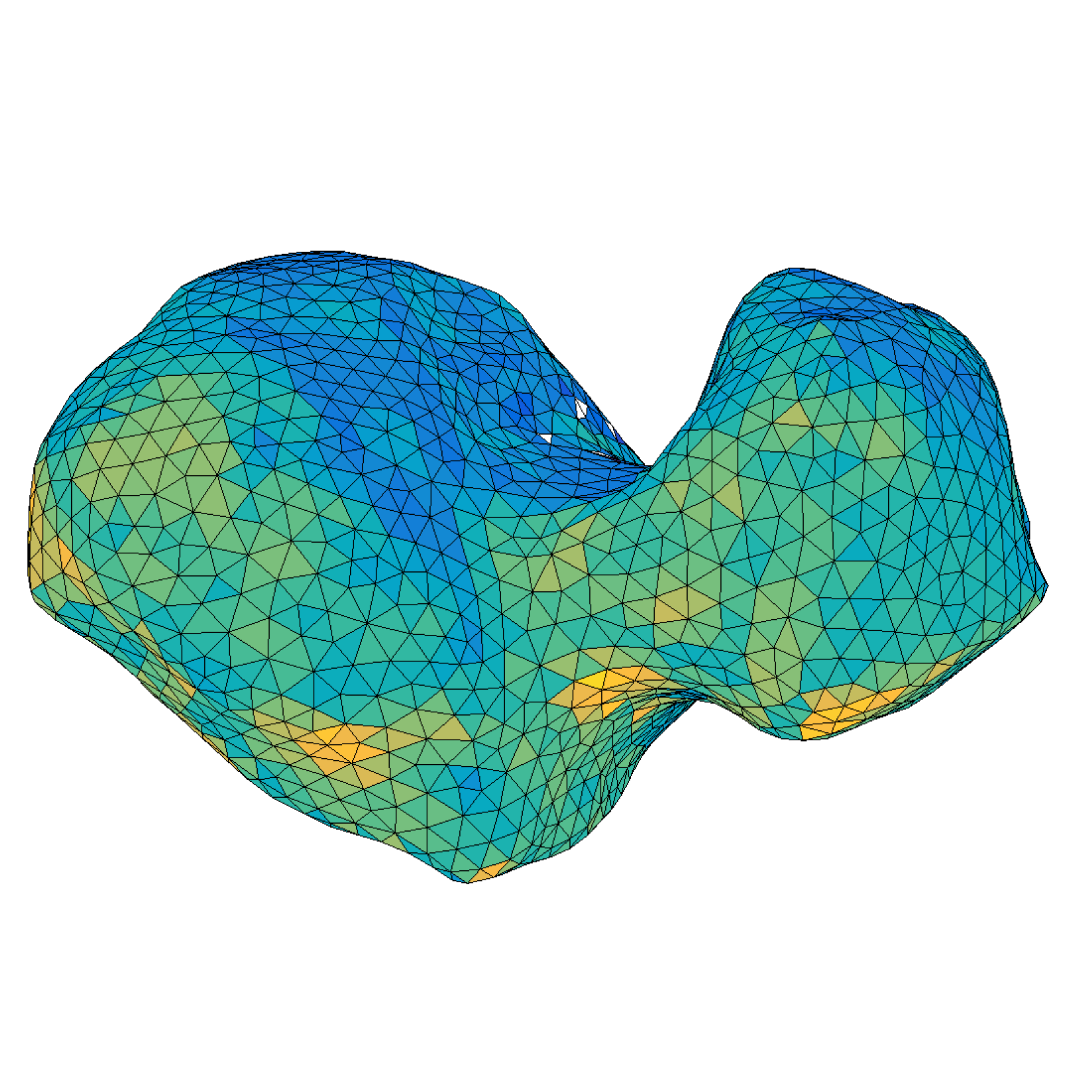}}&
\parbox{0.2\textwidth}{%
\includegraphics[width=0.2\textwidth,draft=false]{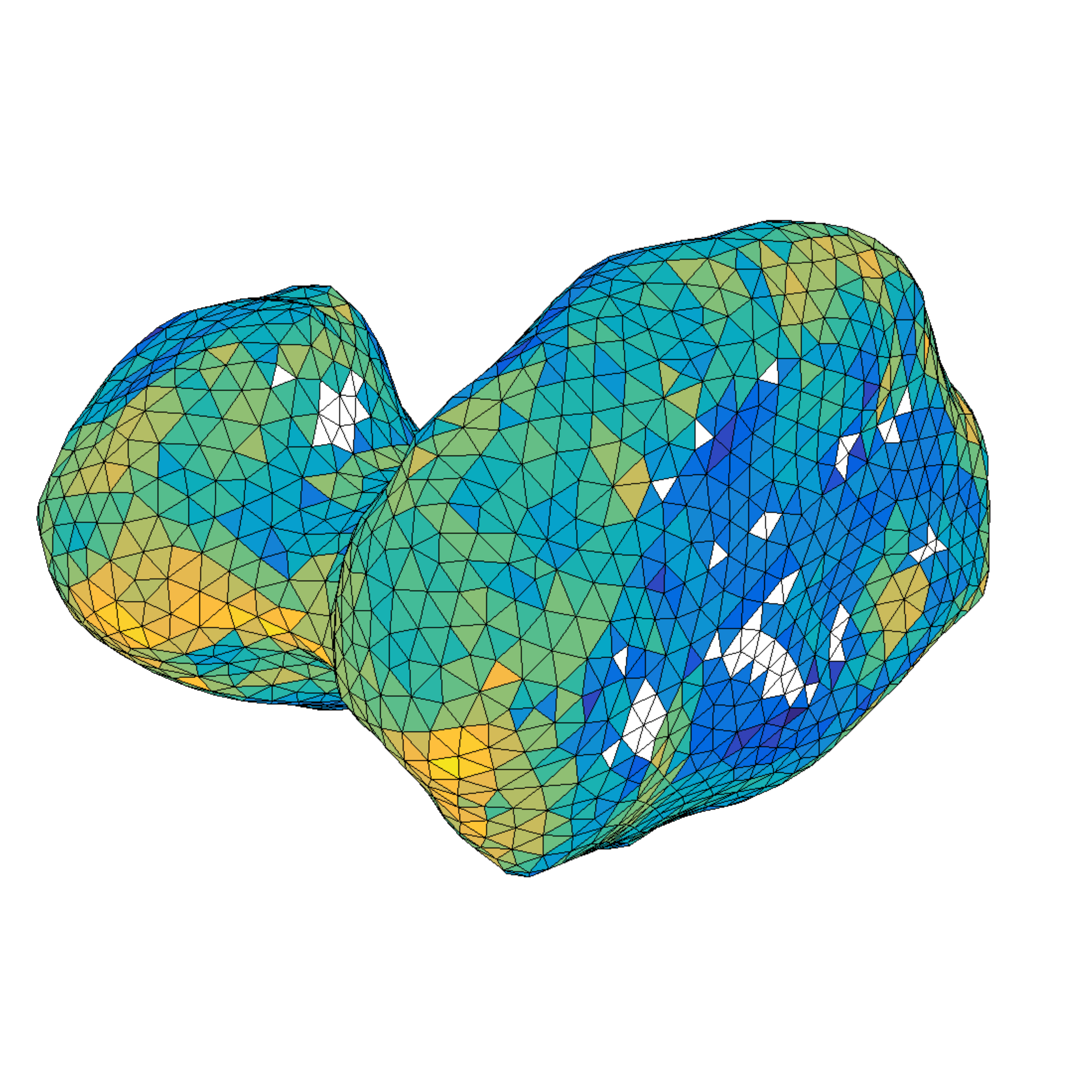}}\\
C&
\parbox{0.2\textwidth}{%
\includegraphics[width=0.2\textwidth,draft=false]{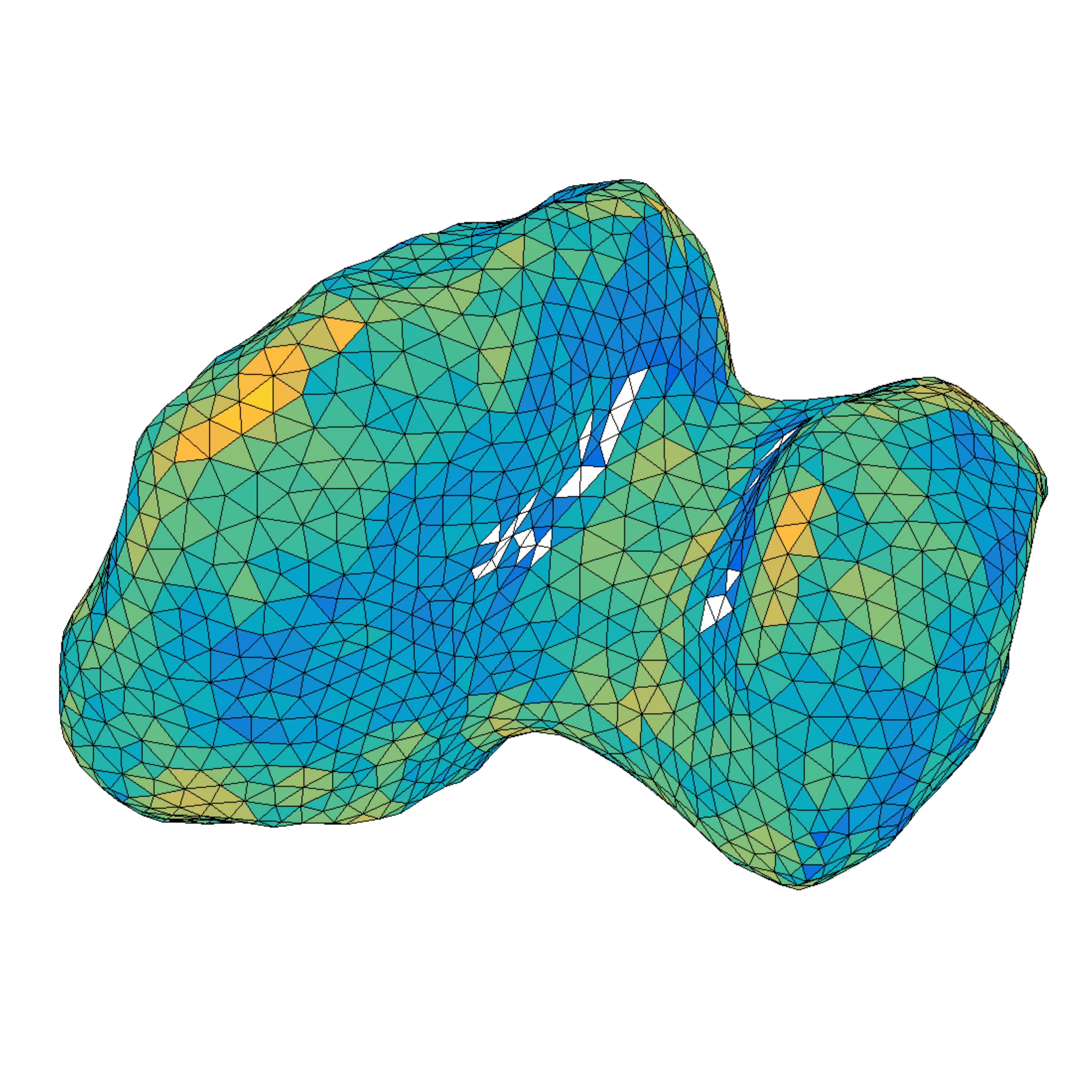}}&
\parbox{0.2\textwidth}{%
\includegraphics[width=0.2\textwidth,draft=false]{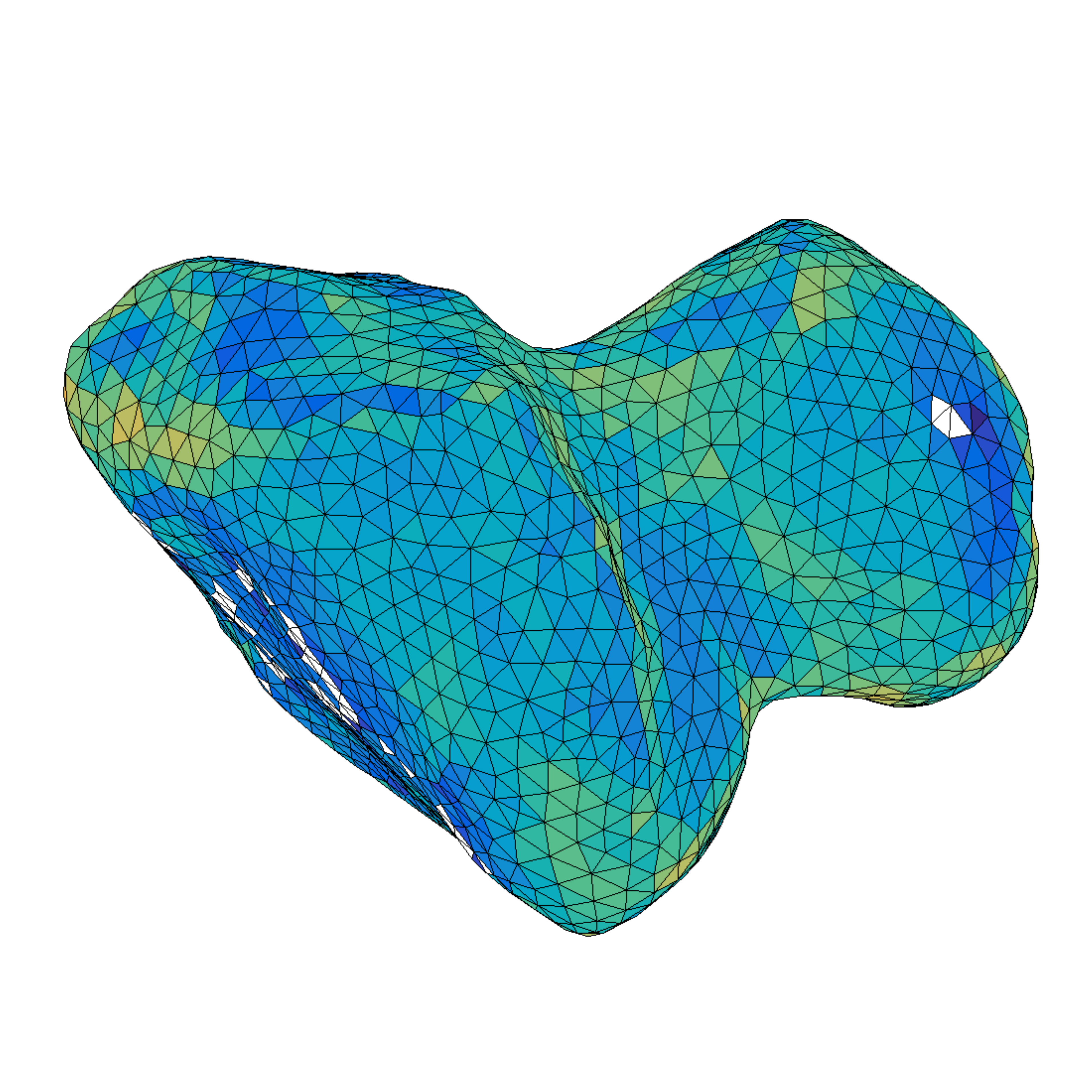}}&
\parbox{0.2\textwidth}{%
\includegraphics[width=0.2\textwidth,draft=false]{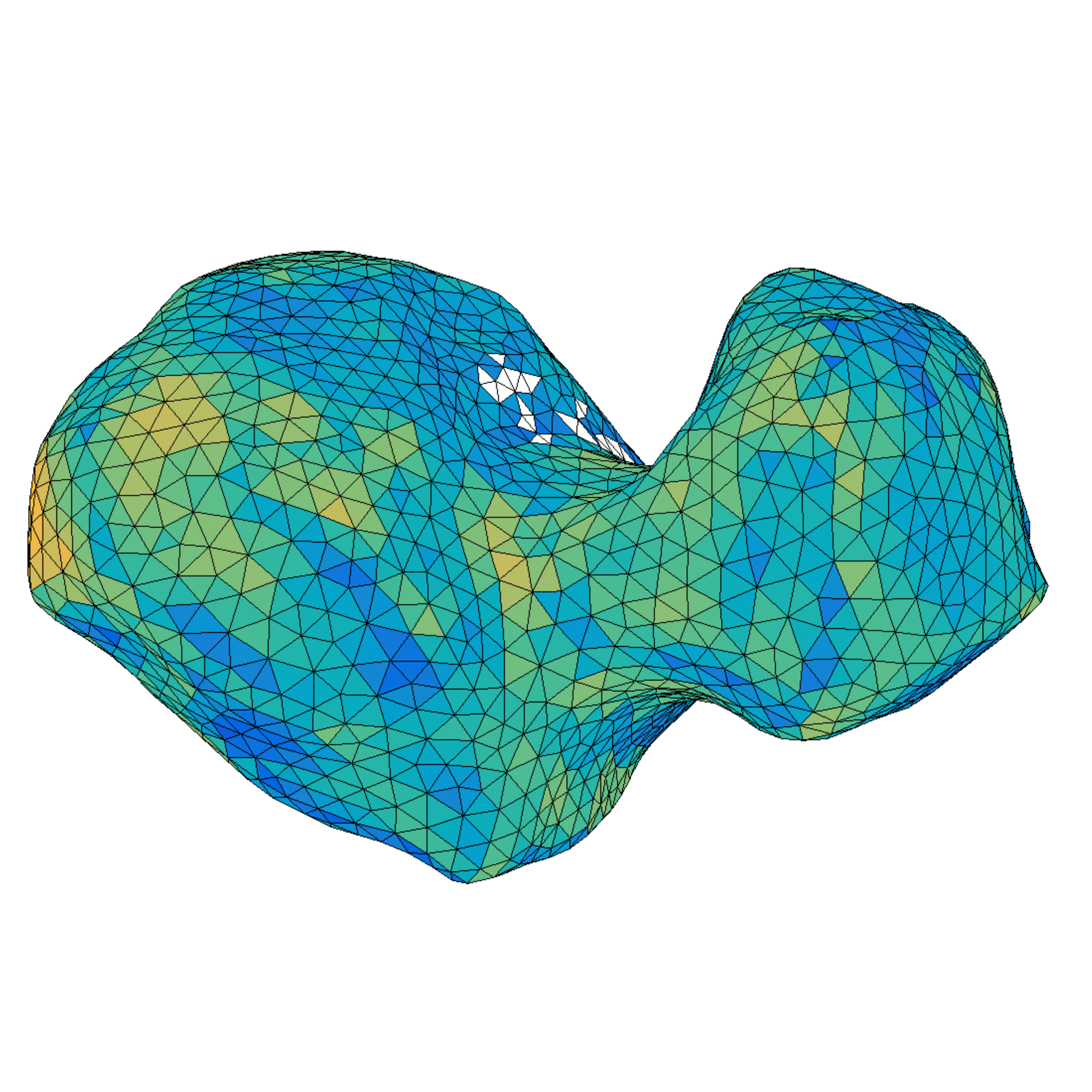}}&
\parbox{0.2\textwidth}{%
\includegraphics[width=0.2\textwidth,draft=false]{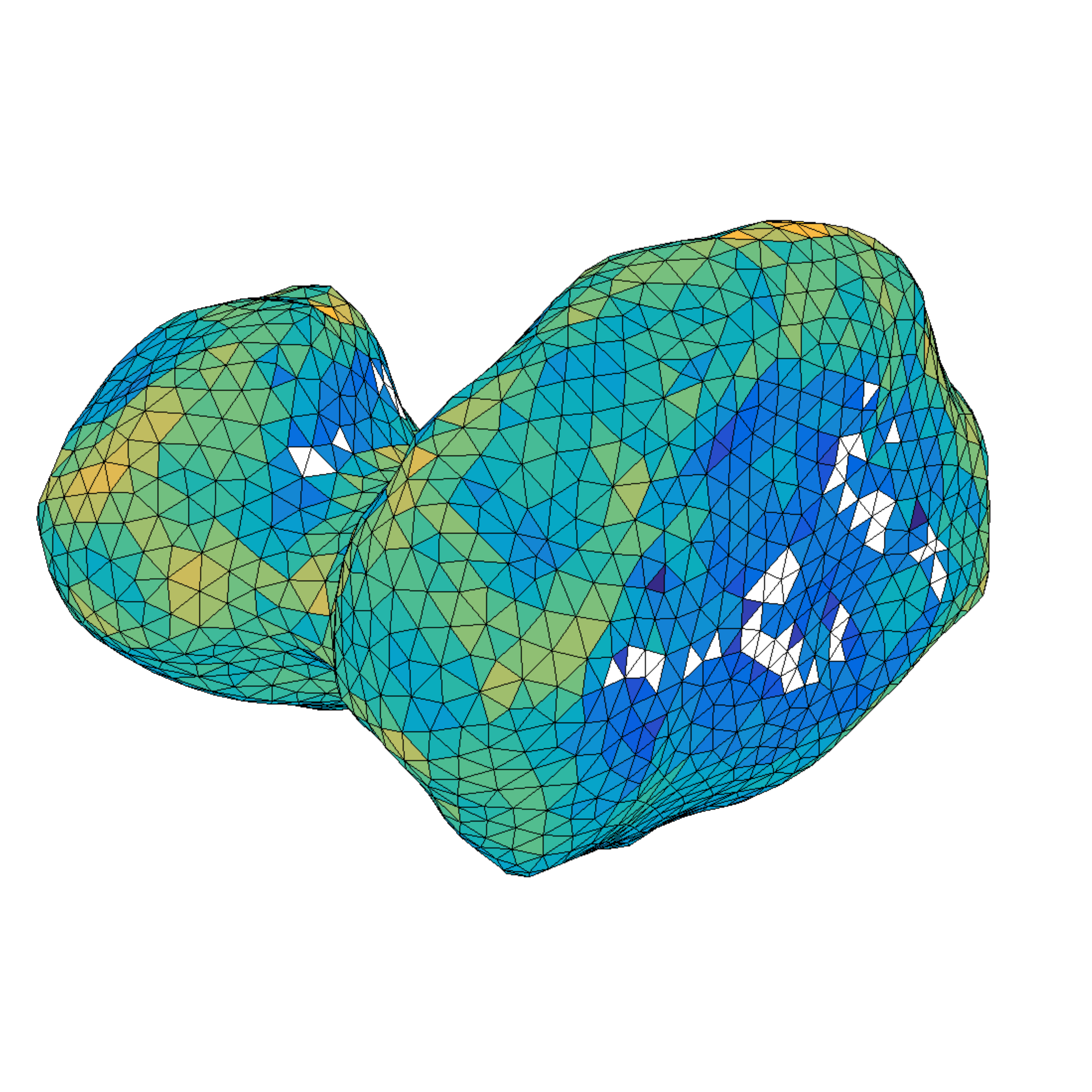}}
\end{tabular}
\caption{Surface emission rate $\dot \rho_{\HiiO,i}$ averaged over the
  intervals $A=(-330,-280)$, $B=(-50,50)$, and $C=(340,390)$ days
  after perihelion.  The colors correspond to the color bars in
  Fig.~\ref{fig:h2oco2} for water and the intervals $A$, $B$, and $C$,
  respectively.}
\label{fig:water3d}
\end{figure*}

Within each subinterval $I_j$ and for every species $s=\HiiO{}$,
$\COii{}$, $\CO{}$, and $\Oii{}$, the gas is emitted constantly in time.
This results in an assimilation of the time-averaged surface emission rates, with a bias toward the local time of observation.
A discussion of density variations due to changing sub-\scraft{}
longitudes follows below.
The surface emission rate for each species $s$ on the surface element
$i=1,...,N_\mathrm{E}$
is given by Eq.~(4) in \cite{Kramer2017}, namely
\[
\dot \rho_{s,i}(t) = \frac{\voutflow{s}}{U_0} q_{s,i}(I_j),
\]
for $t\in I_j$ and $j=1,...,N_\mathrm{I}$, with the speed
$u_{s,0}$ of the outflow velocity into surface normal direction
and the source strength $q_{s,i}$.
The emission rates are expressed in units molecules/m$^2$/s, or
alternatively rescaled to kg/m$^2$/s with the respective molecular
mass.
The parameter $U_0$ denotes the speed ratio between the outflow
velocities along the surface normal $u_{s,0}$ and into the lateral direction.
We treat $U_0$ as an unknown parameter to be determined by a fit and
set the speed into the normal direction as given in
Eqs.~\eqref{eqn:hansen} and \eqref{eqn:vh2o}.
Within the exemplary test interval $(\winbegin,\winend)$~days we have
compared model densities to DFMS/COPS data for different values of
$U_0$, ranging from $U_0=1$ to $U_0=4$.
A larger value $U_0 \ge 4$ exaggerates the density variations at the
sampling points, while a smaller value $U_0 \le 2$ diminishes the
fluctuations.
We have selected $U_0=3$, which gives the best agreement between model and
observations.

The transformation of the DFMS/COPS density data to flux quantities
$\dot \rho_{s,i}(t)$ requires to assign an outflow speed
$\voutflow{s}$ to the density for each interval $I_j$.
At distances $r=10-1000$~km from the nucleus,
\cite{Bockelee-Morvan1987} show that the radiative equilibrium
conditions in the coma lead to speeds around $850$~m/s.
\cite{Lammerzahl1988} measured $800$~m/s at $r=1000-4000$~km for
comet Halley.
DSMC computations by \cite{Tenishev2008} (Fig.~7) and
\cite{Davidsson2010} (Figs.~2,4,5) yield speeds of water of
$900-450$~m/s at heliocentric distances $\helio=1.3-3.5$~au.
For the choice of the speed of water we follow the approach of
\cite{Hansen2016} (Tab.~1, Eq.~7, Fig.~4) and assume a function of
heliocentric distance
\begin{equation}\label{eqn:hansen}
u_{\HiiO,0}(\helio) = u_{\mathrm{Hansen}}(\helio)
\end{equation}
resulting in speeds between $820$~m/s and $560$~m/s.
To facilitate comparisons with other models, we also consider a 
simplified model with a fixed water outflow speed
\begin{equation}\label{eqn:vh2o}
\voutflow{\HiiO} = 755~\mathrm{m/s}.
\end{equation}
If not stated otherwise, the results in this article are based on
Eq.~(\ref{eqn:hansen}).
The speeds of the other species are derived from the water speed
weighted by the square root of the molecular mass ratio with water
\begin{equation*}%\label{eqn:sqrtscale}
\voutflow{s} = u_{\HiiO,0} \sqrt{\mu_\HiiO / \mu_s}.
\end{equation*}

The inverse model for each of the time intervals consists of a
fitting process to determine all surface emission
rates of the four major volatiles $\HiiO$, $\COii$, $\CO$, and $\Oii$.
A typical, species-resolved density reconstruction within four
intervals is shown in Fig.~\ref{fig:copsdfms_detail}(c).
Similar to \cite{Bieler2015} we observe periodic density
variations (approximately two maxima per orbit around the nucleus) 
in the DFMS/COPS data and also for our modelled densities at the \scraft{}
positions.
The Rosetta orbit mostly follows a terminator geometry, leading to
preferential observations at morning/evening phase angles.
Because our model assimilates diurnally averaged production rates
within each subinterval $I_j$, changing illumination conditions are
not resolved.
The model fits are interpreted as diurnally averaged production rates
of localized gas sources which reflect fluctuations due to changes of
the sub-\scraft{} position.
This interpretation is supported by the consistent retrieval of
activity spots across the entire mission from independently processed
COPS/DFMS data sets taken months apart.

The model performance depends on the DFMS/COPS data distribution in
time and space.
In each interval $I_j$ the fit performance is quantified by the
relative $l2$ error norm of the difference of predicted and measured
densities at times $T_{4\mathrm{h}}\cap I_j$.
All errors are in the range $0.04-0.50$, with an average value $0.16$.
Possible error sources are temporal changes in surface activity
or deviations from the collisionless gas model.
The construction of the global emission map depends on the surface
coverage of the nucleus by the \scraft{} within each interval $I_j$.
Even a limited coverage yields a subset of surface elements with
known gas emission rates.
We assign the source strength $q_{s,i}(I_j)$ for an uncovered element $E_i$  either from a minimum, a linear, or a maximum estimate.
Based on the neigboring values $l=q_{s,i}(I_{j-1})$,
$r=q_{s,i}(I_{j+1})$, the minimum estimation sets $q_{s,i}(I_j)=0$,
the linear estimation sets $q_{s,i}(I_j)$ to the average of $l$
and $r$, and the maximum estimation sets $q_{s,i}(I_j)=\max(l,r)$.
The production rates in the article are based on the linear estimate,
the uncertainty values are based on the minimum and maximum estimates.
The minimum estimation provides a strict lower limit, while the maximum
estimation provides only a heuristic upper limit since local maxima could be dismissed.
Thus, the \scraft{} coverage errors could lead to an underestimation
of production rates.
The productions rates along with the minimum and maximum estimates are
shown in Fig.~\ref{fig:qtimespecies}.

\section{Global gas production}\label{sec:qtotal}

\begin{table*}
\begin{tabular}{lcccl}
$s$ & $\mission{P}{s}{}$~[molecules] & $\mission{P}{s}{}$~[kg] &
 $\mission{P}{s}{,S} / \mission{P}{s}{,N}$ &
 $\max \mission{Q}{s}{}$~[kg/s] \\\hline
$\HiiO$ & $1.6\pm 0.5 \times 10^{35}$ &
$4.8\pm 1.5 \times 10^9$ & $2.0$ &
$6.1 \pm 0.3\times 10^2$ \\
$\COii$ & $9.5\pm 3.8 \times 10^{33}$ &
$7.0\pm 2.8 \times 10^8$ & $4.9$ &
$8.6 \pm 0.4\times 10$ \\
$\CO$   & $3.6\pm 1.1 \times 10^{33}$ &
$1.7\pm 0.5 \times 10^8$ & $1.7$ &
$2.0 \pm 0.1\times 10$ \\
$\Oii$  & $2.6\pm 0.8 \times 10^{33}$ &
$1.4\pm 0.4 \times 10^8$ & $1.8$ &
$1.4 \pm 0.7\times 10$
\end{tabular}
\caption{Integrated production $\mission{P}{s}{}$ from
  Eq.~\eqref{eqn:totalsum} for the species $s$, relative rates
  $\mission{P}{s}{,S} / \mission{P}{s}{,N}$ between production rates
  resolved by north (N) and south (S) emission location,
  peak production rates $\max \mission{Q}{s}{}$.
}
\label{tab:qint}
\end{table*}

The spatially integrated production rates $\mission{Q}{s}{}(t)$ follow
directly from the spatially and temporally resolved surface rates
$\dot \rho_{s,i}(t)$ by summing over all $E_i$ shape elements
\[
\mission{Q}{s}{}(t) = \sum_{i=1}^{N_\mathrm{I}} \dot \rho_{s,i}(t) \; {\rm area}(E_i)
\]
for the gas species $s$.
The integrated productions $\mission{P}{s}{}$ in space and time during
the 2015 apparition
are obtained by
\begin{equation}\label{eqn:totalsum}
\mission{P}{s}{} = \int_{\misbegin~\mathrm{days}}^{\misend~\mathrm{days}}
 Q_s(t)\, \mathrm{d} t.
\end{equation}
Similar to $\dot \rho_{s,i}(t)$, all production quantities
depend on the molecular speeds $\voutflow{s}$, see
Eqs.~\eqref{eqn:hansen},\eqref{eqn:vh2o}.

For an outflow speed depending on the heliocentric distance
$\helio$ (Eq.~\eqref{eqn:hansen}), Fig.~\ref{fig:qtimespecies} shows
productions rates as a function of $\helio$ and of time for all
species $\HiiO$, $\COii$, $\CO$, and $\Oii$.
Table~\ref{tab:qint} lists the integrated productions
$\mission{P}{s}{}$ and the peak productions $\max \mission{Q}{s}{}$.
The alternative model with an overall constant outflow speed 
Eq.~\eqref{eqn:vh2o} leads to similar integrated production rates.
The peak gas production of $2.2\pm 0.1\times 10^{28}$~molecules/s 
($730\pm 30$~kg/s) is reached in the interval $I=(17,28)$~days
after perihelion and is clearly dominated by $\HiiO$, whereas $\COii$
contributes with only one tenth of the water mass production.
Compared to that, the model with constant speed (Eq.~\eqref{eqn:vh2o}) 
results in a reduced peak production of $2.1\pm
0.1\times 10^{28}$~molecules/s ($690\pm 30$~kg/s).
For water, the peak production yields
$\max\mission{Q}{\HiiO}{}$ is $2.0\pm 0.1\times 10^{28}$~molecules/s
and the integrated production $\mission{P}{\HiiO}{}$ for one orbit
yields $4.8\pm1.5\times 10^9$~kg.
Assuming the same outflow speed (Eq.~\eqref{eqn:hansen}),
\cite{Hansen2016} derive from COPS data a peak water production of
$3.5\pm 0.5 \times 10^{28}$~molecules/s $18-22$~days after perihelion.
One possible reason for the higher value given by \cite{Hansen2016}
might be the different interval lengths used for averaging the data
(four days compared to eleven days in our case).
The integrated water production of $6.4\times 10^{9}$~kg per orbit
from \cite{Hansen2016} is in better agreement with our estimate.
From the MIRO analysis \cite{Marshall2017} obtain a highest water
emission of $1.42\pm 0.51 \times10^{28}$~molecules/s 16~days after
perihelion.
Their integrated water production of $2.4\pm 1.1\times10^9$~kg for
the apparition 2015 is half of our value.
One possible cause could be a distributed source of e.g. icy grains
that evaporate before reaching Rosetta where they are measured by
ROSINA but do not contribute close to the nucleus to the measurements
of MIRO.
Another approach from \cite{Shinnaka2017} is to consider the hydrogen
Lyman $\alpha$ emissions.
25~days after perihelion they obtain a water production rate of
$1.46\pm 0.47\times 10^{28}$~molecules/s.
The $\HiiO$ productions based on MIRO and Lyman $\alpha$ data are not
peak values and thus correspond to our lower estimate.

The orbital losses allow us to constrain the dust-to-gas ratio of
\shortcomet{}.
The total gas loss $P_{\mathrm{gas}}$ is considered to be the
contributions from $\HiiO$, $\COii$, $\CO$, and $\Oii$ and further
$5\%$ volatile and massive species like $\mathrm{CS_2}$,
$\mathrm{H_2S}$, $\mathrm{SO_2}$, $\mathrm{C_2H_6}$, see
\cite{LeRoy2015} and \cite{Calmonte2016}.
This yields $P_{\mathrm{gas}} = 1.05 \cdot(P_{\HiiO} + P_{\COii}
+P_{\CO} +P_{\Oii}) = 6.1\pm 1.9\times 10^9$~kg
and corresponds to $1/1600$ of the total mass of $M_{\shortcomet{}} =
9.9778\pm 0.004\times 10^{12}$~kg from \cite{Godard2017}.
Considering the mass for October 2014 in \cite{Godard2015}, their
estimation for the total mass loss is $P_{\mathrm{dust}+\mathrm{gas}}
= 9\pm 6\times 10^9$~kg including a significant uncertainty.
This uncertainty propagates to the dust-to-gas ratio of the emitted
material, which we estimate to be $(P_{\mathrm{dust}+\mathrm{gas}} -
P_{\mathrm{gas}})/P_{\mathrm{gas}} = 0.5^{+1.1}_{-0.5}$.
This value presents a lower limit for the dust-to-gas ratio.
The escaping material may still contain volatiles which affect the
dust-to-gas ratio, see e.g. \cite{Dekeyser2017} and
\cite{Altwegg2016b}.
In addition, the dust-to-gas ratio may differ from the dust-to-ice
ratio in the nucleus as backfall of dry or almost dry dust would
contribute to the amount of dust ejected, but would not lead to mass
loss of the nucleus.

The sufficient temporal coverage of DFMS/COPS data allows us to
integrate the production per orbit by summing all
interval contributions, see Eq.~\eqref{eqn:totalsum}.
Another possibility sometimes used in the literature
is to approximate the integral from the power law fit $\helio^\alpha$.
Fig.~\ref{fig:qtimelinear} shows that the production rate
$\mission{Q}{\HiiO}{}$ follows power laws with exponents
$\helio^{-7}$ and $\helio^{-6.5}$ for the inbound and
outbound orbits, respectively.
The exponents given by \cite{Hansen2016} ($-5.1\pm0.05$ and
$-7.15\pm0.08$) and \cite{Shinnaka2017} ($-6.0\pm0.46$ and
$-5.22\pm0.41$) are in a similar range.
The data analysis of \cite{Marshall2017} yields considerably 
lower exponents ($-3.8\pm0.2$ inbound, $-4.3\pm0.2$ outbound).
This is one consequence of the smaller peak production rates
derived from MIRO versus ROSINA as discussed above in the context of
the peak production.
Although not as steep as for $\HiiO$, the $\Oii$ curves are
fitted by exponents of $-5.5$ and $-6$.
The inbound production of $\COii$ and $\CO$ is not well reproduced by
a power law, since $150$~days before perihelion and even earlier the 
production rate stagnates.
Outbound, the $\COii$ production drops down with $\helio^{-4.5}$,
slower than for $\HiiO$.
This difference leads to a crossover from a water dominated coma to a
carbon dioxide dominated one at $2.75$~au ($250$~days after
perihelion).
$\CO$ partially resembles the $\COii$ trend with a similar exponent
$\helio^{-6.0}$.

Fig.~\ref{fig:qtimelinear} and Table~\ref{tab:qint} show production
contributions separated for the Northern (N) and Southern (S)
hemispheres.
All species are released in higher quantities from the southern
hemisphere compared to the northern one.
This is caused by the stronger illumination of the southern latitudes
during perihelion, with summer solstice occurring only 23 days after
perihelion.
The asymmetric mass production ratios
$\mission{P}{s}{,S}/\mission{P}{s}{,N}$ for $\HiiO$, $\CO$, and $\Oii$
range between $1.7:1$ to $2.0:1$.
In contrast to that, the S/N ratio for $\COii$ becomes $4.9:1$.
This indicates a predominant $\COii$ production from
southern sources.
In agreement with the southwards shifted integrated productions, the
ratios $\mission{Q}{s}{,S}(t)/\mission{Q}{s}{,N}(t)$ around perihelion
are close to the S/N ratios in Table~\ref{tab:qint} for
$\mission{P}{s}{}$.
For $\CO$, the S/N ratio remains elevated also on the outbound
cometary orbit after perihelion and for $\COii$ at almost all times.
For $\COii$, only the first interval is an exception, where the
sub-\scraft{} latitude leads to a poor southern coverage.

\begin{figure}
\includegraphics[width=0.45\textwidth,draft=false]{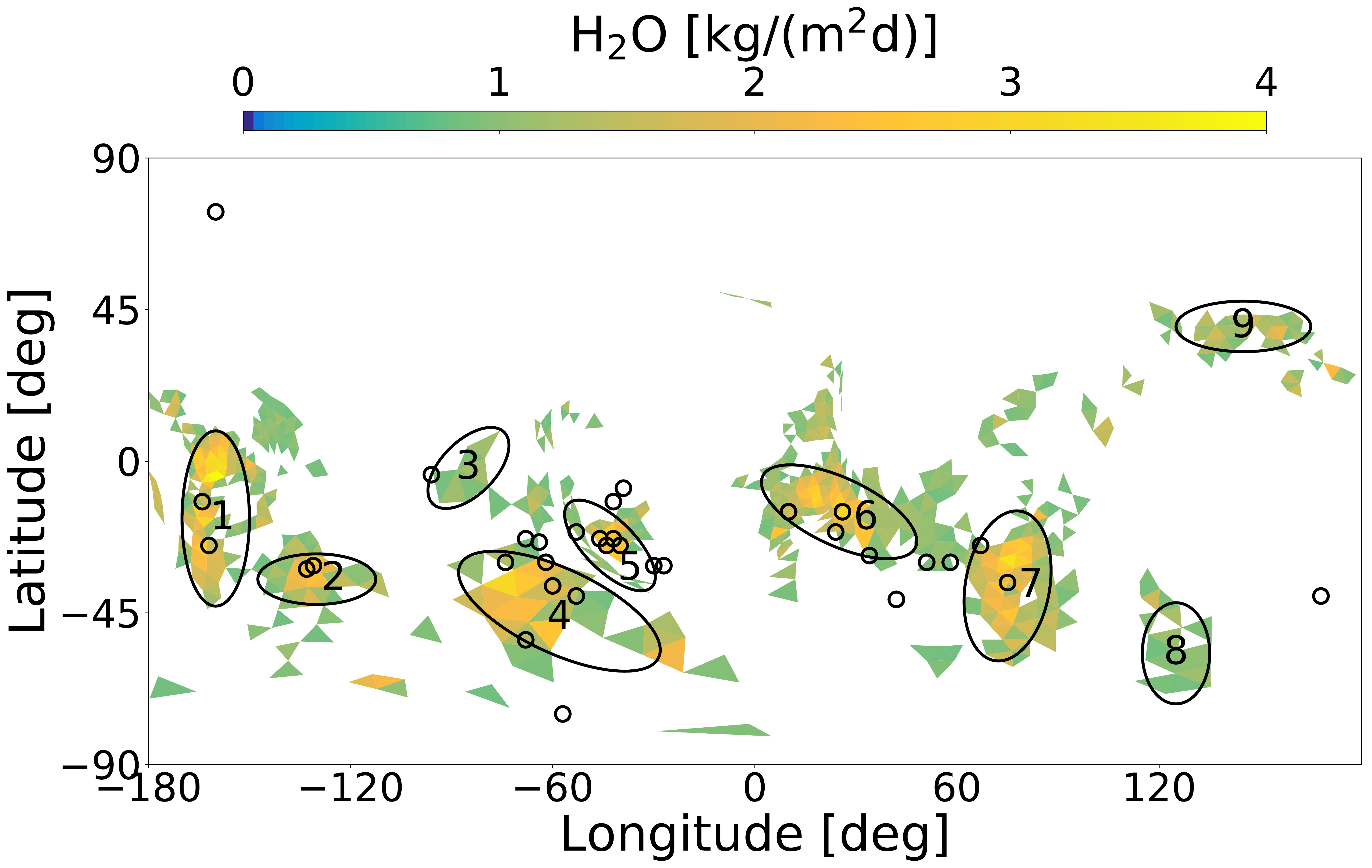}
\caption{Surface emission rates $\dot \rho_{\HiiO,i}$ in the time
  interval $B=(-50,50)$~days on the most active surface elements,
  contributing to 50\% of the total emission; \nine{} $\HiiO$ activity
  areas are marked by ovals.  The circles show the positions of
  reported short living outbursts by \protect\cite{Vincent2016}.  }
\label{fig:focusareas}
\end{figure}

\begin{figure*}%
\begin{tabular}{lcc}
A&
\parbox{0.45\textwidth}{%
\includegraphics[width=0.45\textwidth,draft=false]{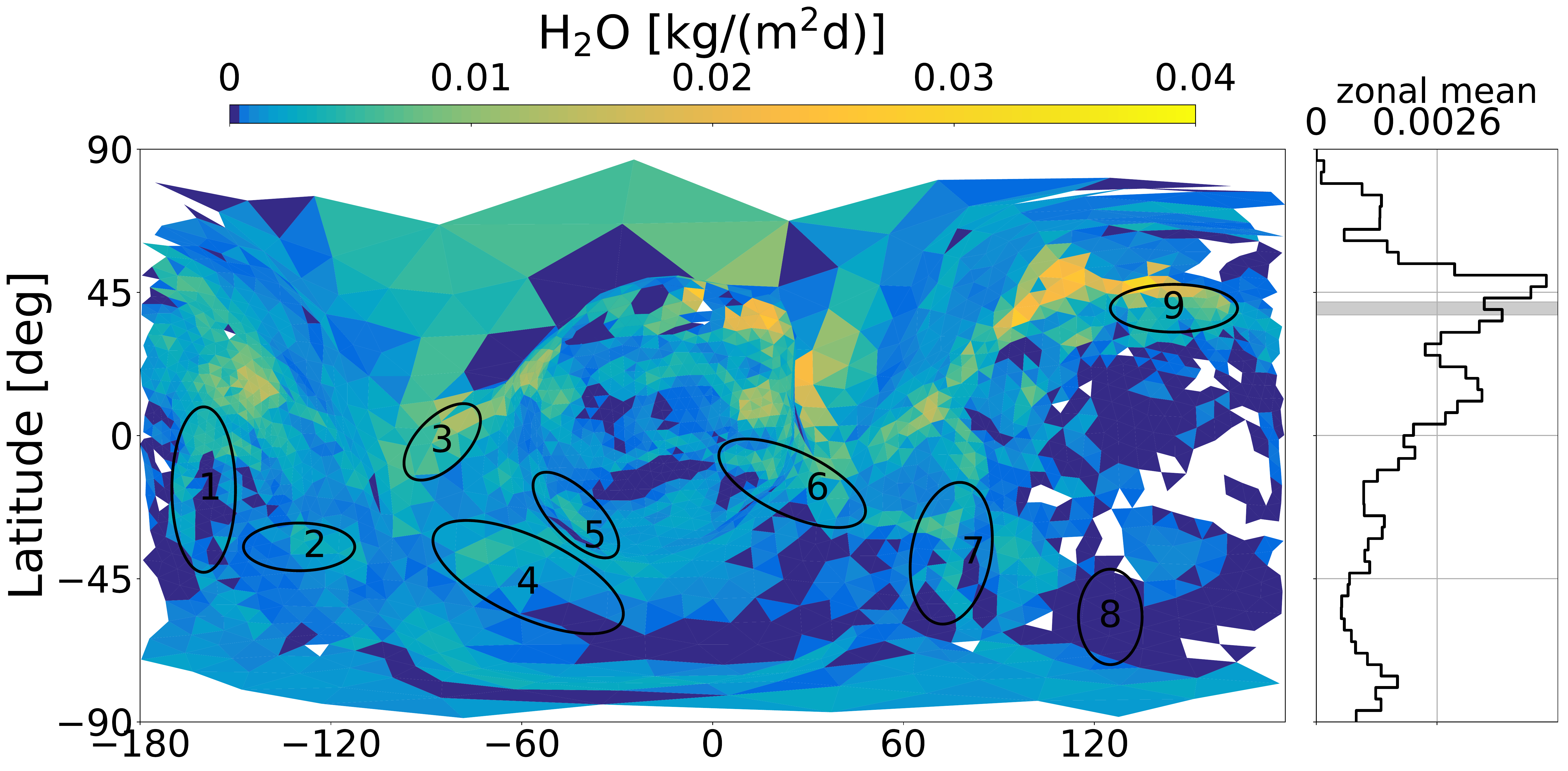}}&
\parbox{0.45\textwidth}{%
\includegraphics[width=0.45\textwidth,draft=false]{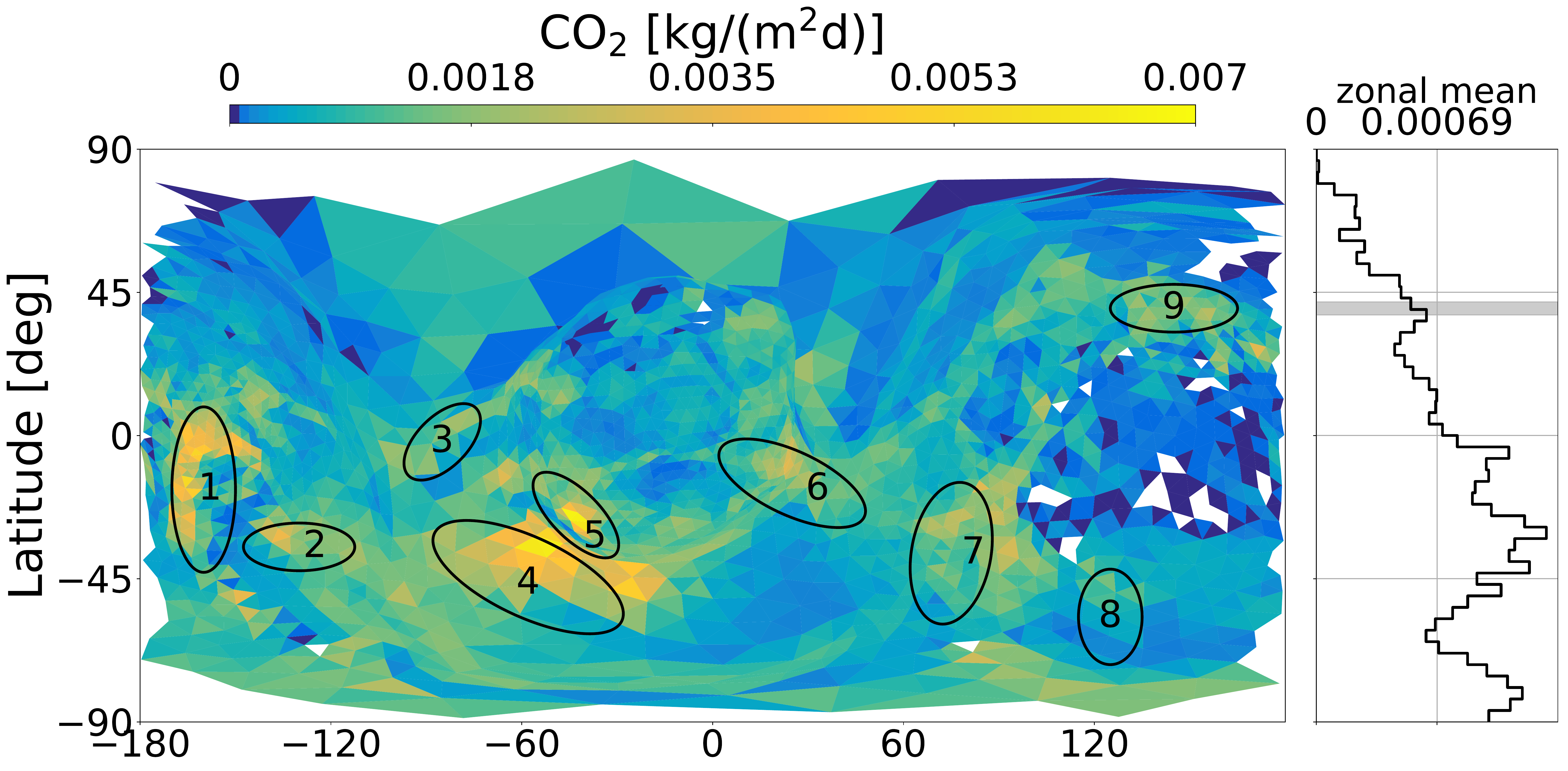}}\\
B&
\parbox{0.45\textwidth}{%
\includegraphics[width=0.45\textwidth,draft=false]{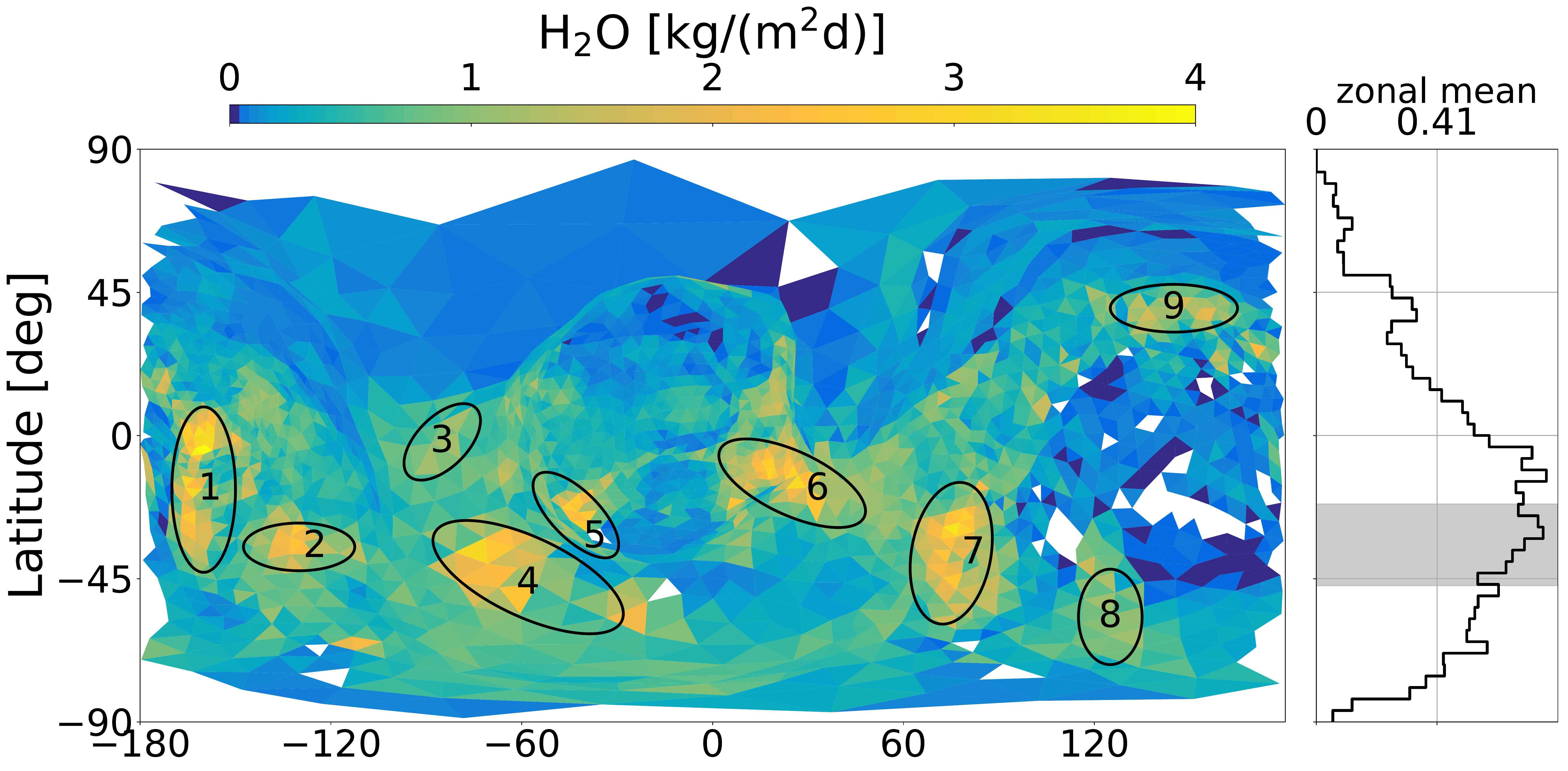}}&
\parbox{0.45\textwidth}{%
\includegraphics[width=0.45\textwidth,draft=false]{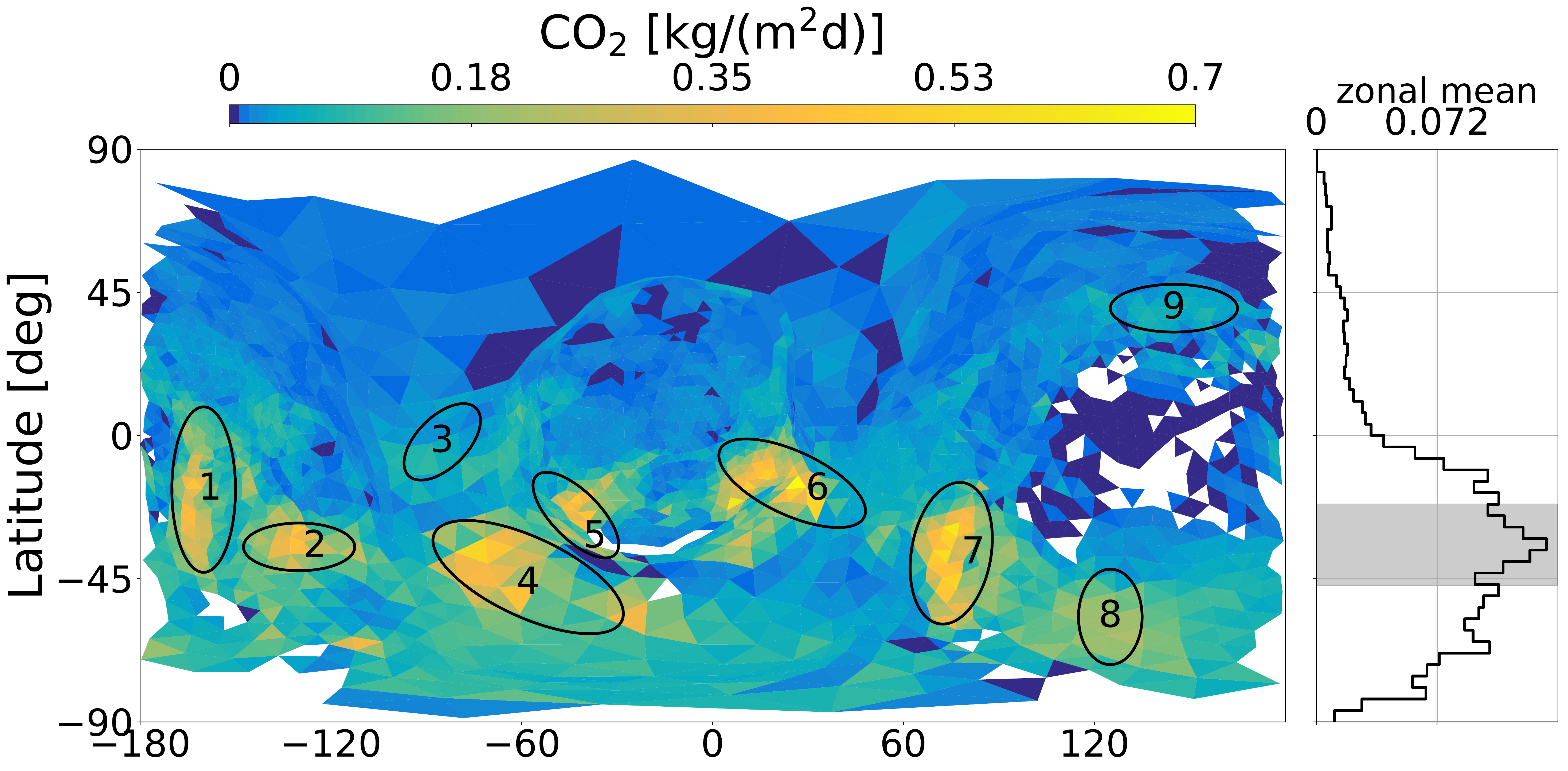}}\\
C&
\parbox{0.45\textwidth}{%
\includegraphics[width=0.45\textwidth,draft=false]{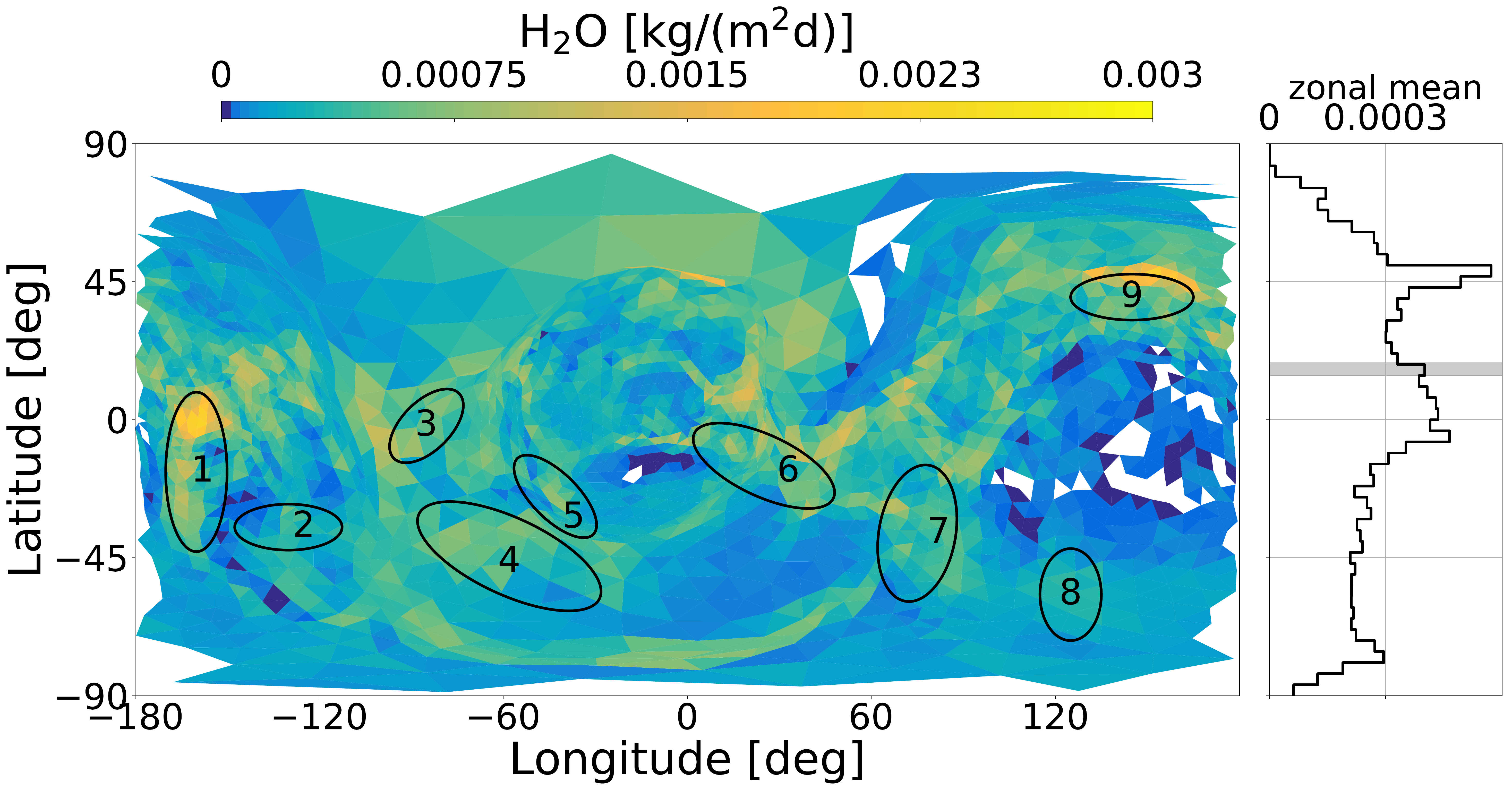}}&
\parbox{0.45\textwidth}{%
\includegraphics[width=0.45\textwidth,draft=false]{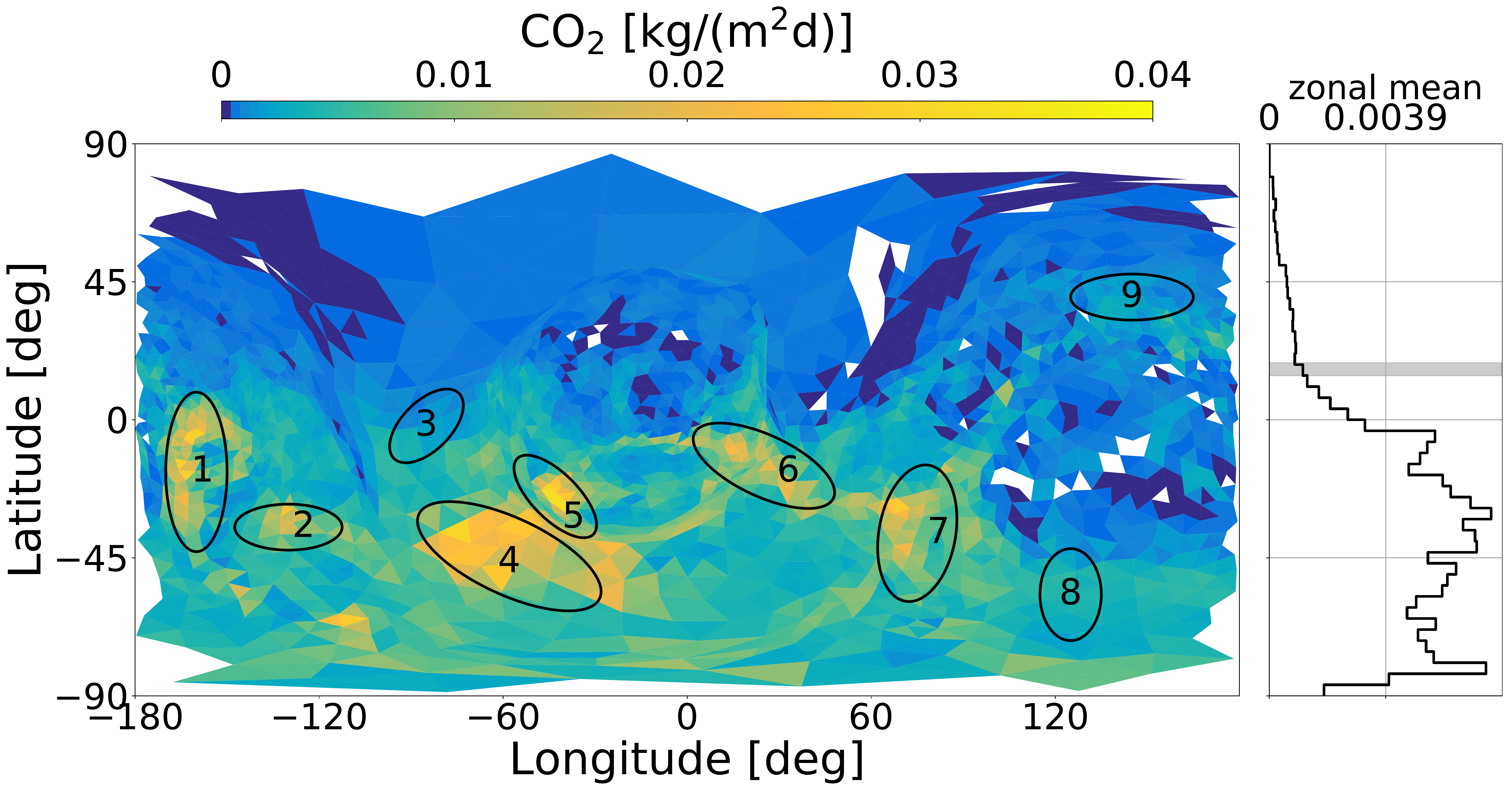}}
\end{tabular}
\caption{Surface emission rates $\dot \rho_{\HiiO,i}$ and $\dot
  \rho_{\COii,i}$ in the intervals $A=(-330,-280)$, $B=(-50,50)$, and
  $C=(340,390)$ days after perihelion.  The side panels show the
  longitudinally averaged rate (zonal mean), the grey bar indicates
  the sub-solar latitude.}
\label{fig:h2oco2}
\end{figure*}

\begin{figure*}
\begin{tabular}{lcc}
A&
\parbox{0.45\textwidth}{%
\includegraphics[width=0.45\textwidth,draft=false]{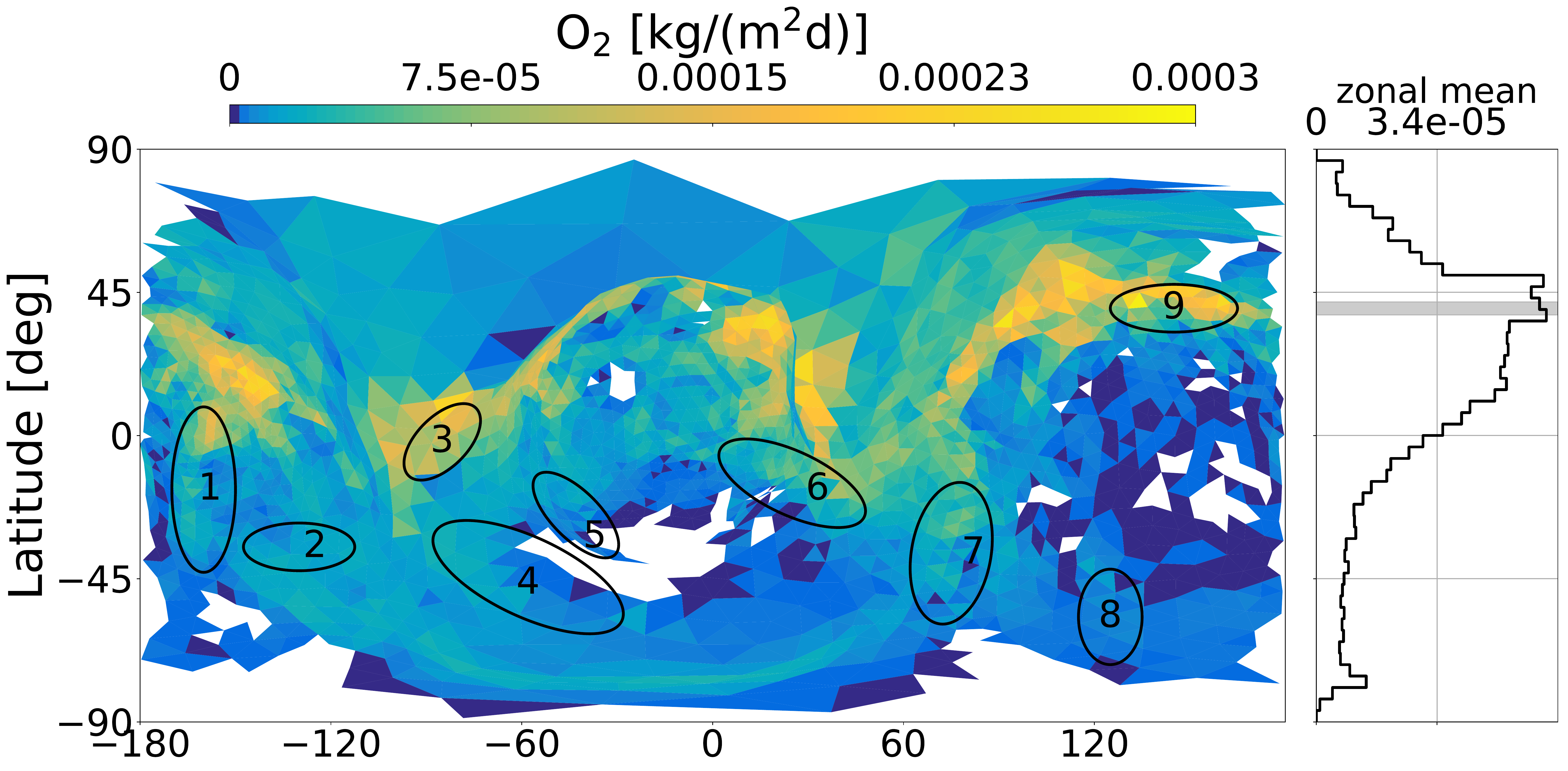}}&
\parbox{0.45\textwidth}{%
\includegraphics[width=0.45\textwidth,draft=false]{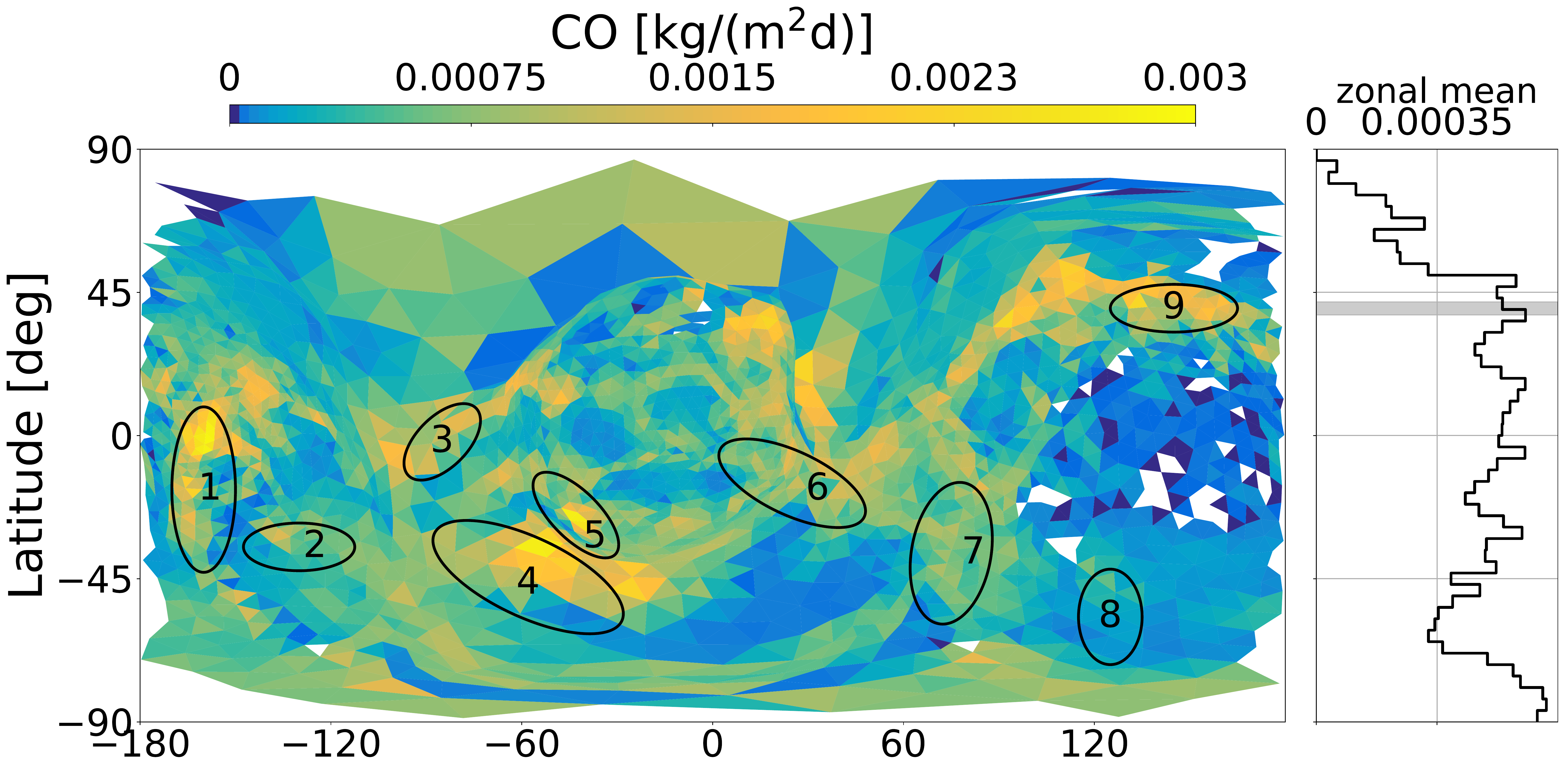}}\\
B&
\parbox{0.45\textwidth}{%
\includegraphics[width=0.45\textwidth,draft=false]{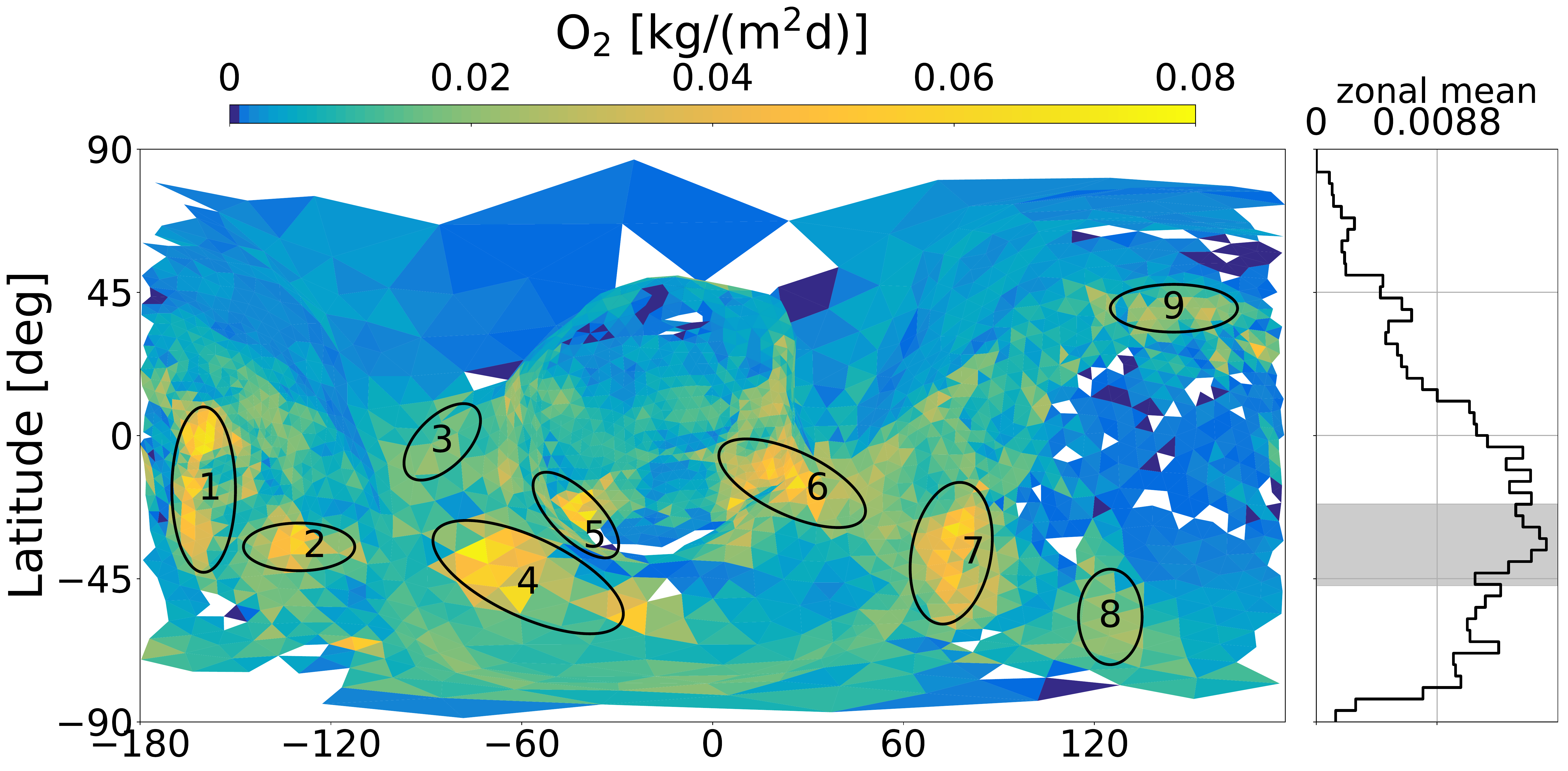}}&
\parbox{0.45\textwidth}{%
\includegraphics[width=0.45\textwidth,draft=false]{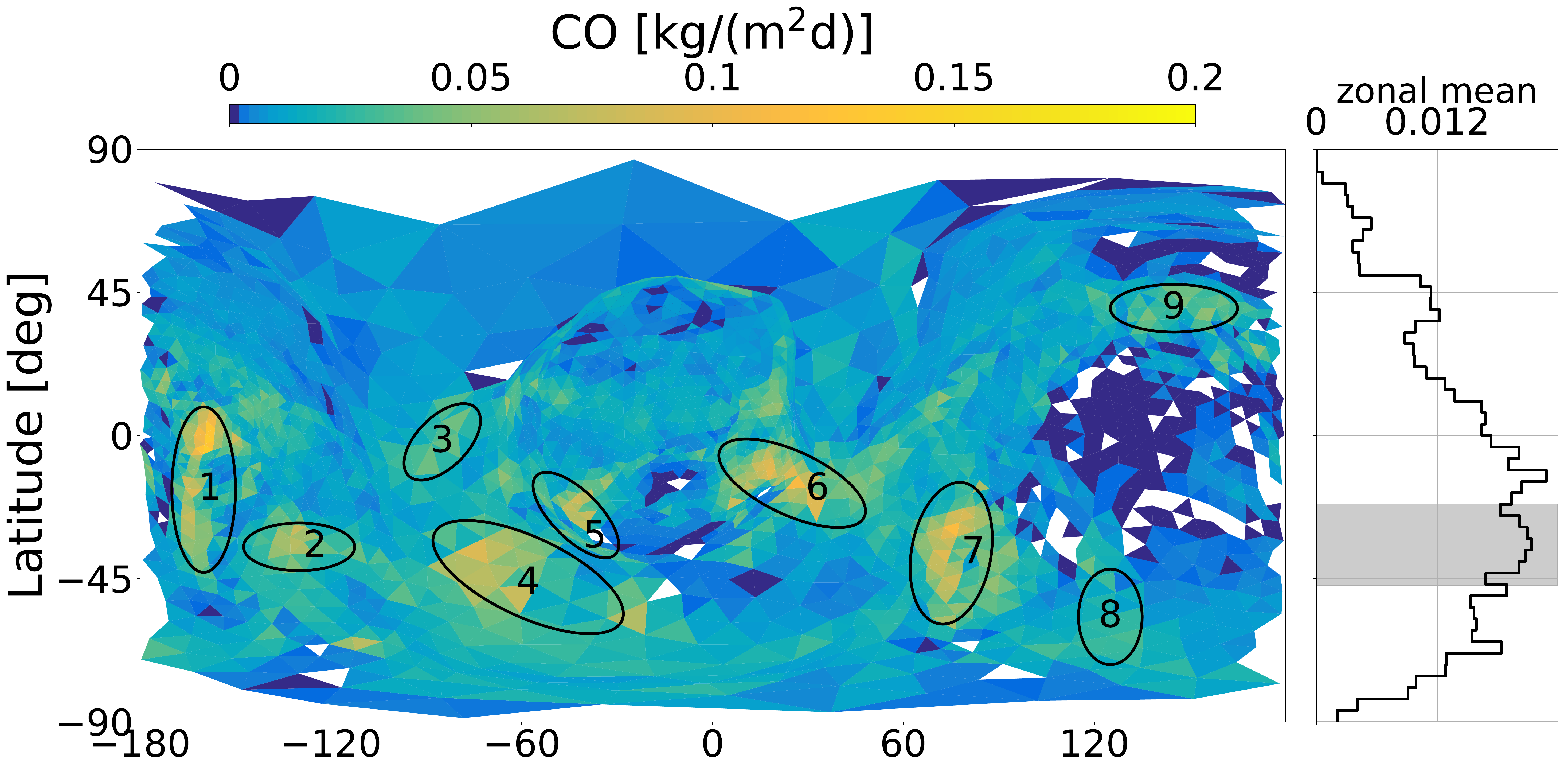}}\\
C&
\parbox{0.45\textwidth}{%
\includegraphics[width=0.45\textwidth,draft=false]{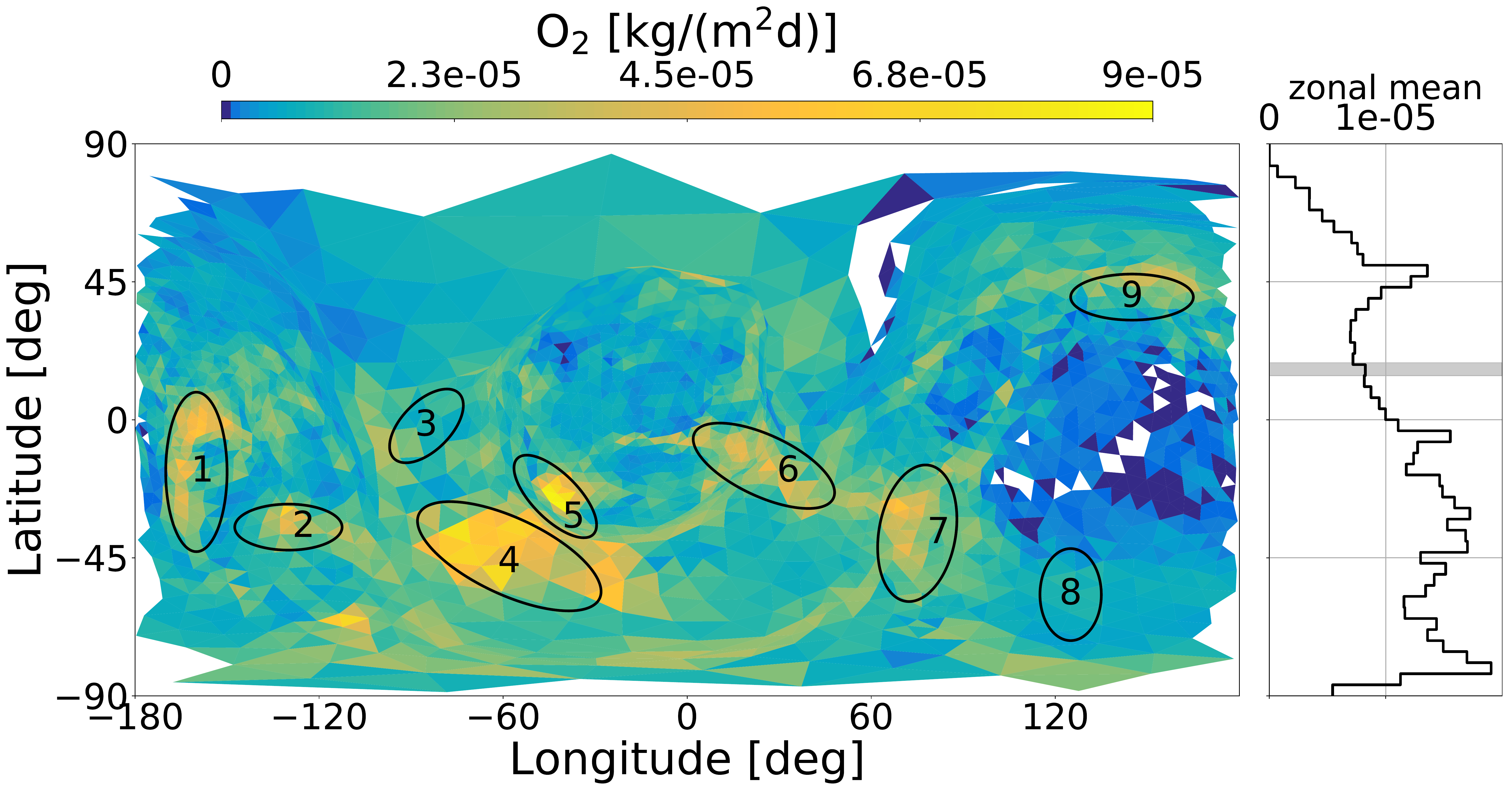}}&
\parbox{0.45\textwidth}{%
\includegraphics[width=0.45\textwidth,draft=false]{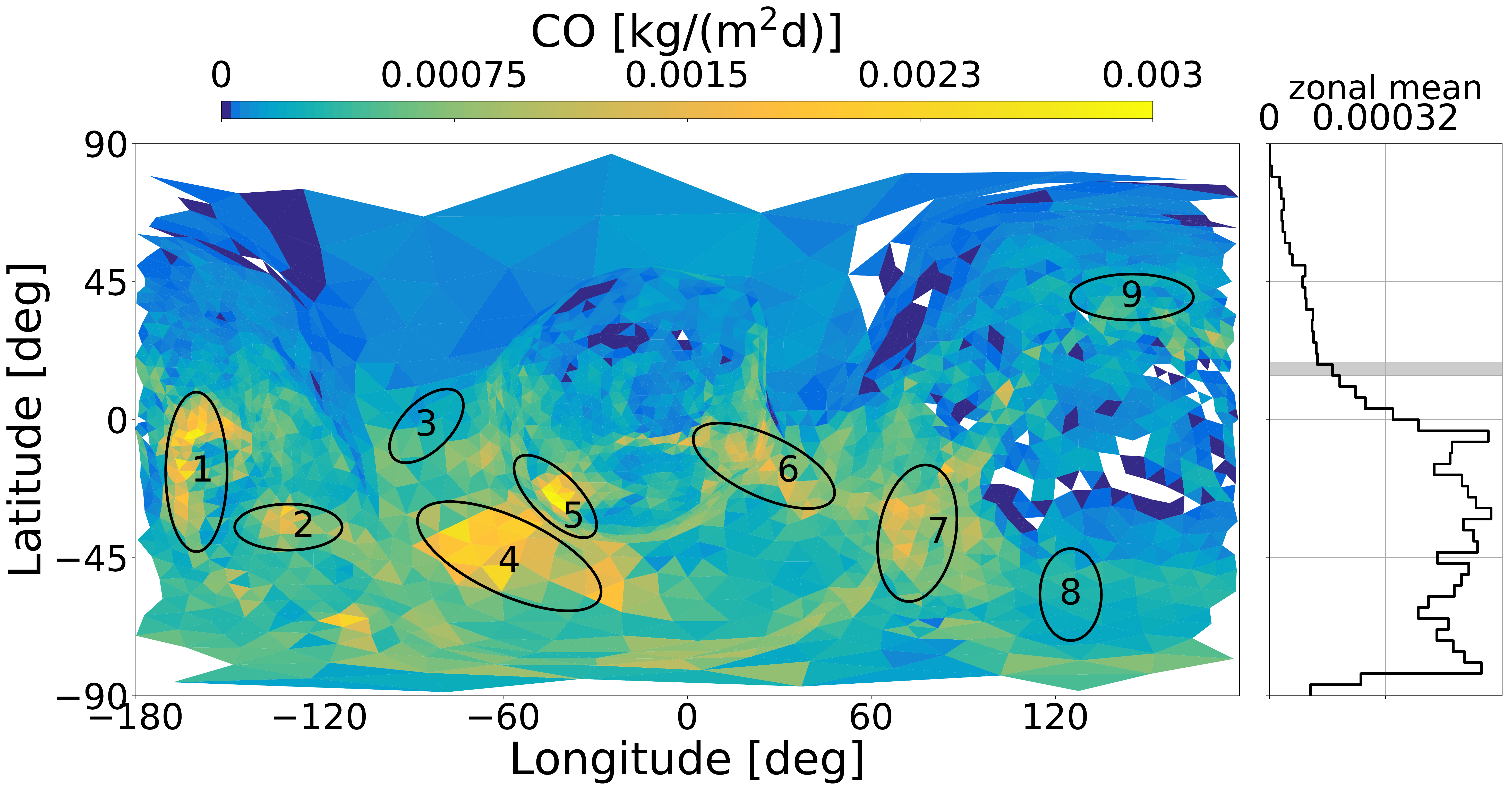}}
\end{tabular}
\caption{Surface emission rates $\dot \rho_{\Oii,i}$ and $\dot
  \rho_{\CO,i}$ in the intervals $A=(-330,-280)$, $B=(-50,50)$, and
  $C=(340,390)$~days after perihelion.  The side panels show the
  longitudinally averaged rate (zonal mean), the grey bar indicates
  the sub-solar latitude.  }
\label{fig:o2co}
\end{figure*}

\section{Localized surface sources}\label{sec:surface}

It has been recognized, see e.g. \cite{Bieler2015}, that a homogeneous
distriubtion of the activity cannot explain the coma gas distribution.
Consequently advanced models use different heterogeneous distributions
of active areas.
E.g. \cite{Fougere2016a} use an inverse approach for spherical
harmonics in the neck region to introduce heterogeneity,
\cite{Marschall2017} use specific surface morphology (cliffs, plains)
to attribute activities to different areas.
Our inverse model allows one to trace back in situ DFMS/COPS
measurements in the coma to localized emission rates.
It incorporates the complex shape of the nucleus with two lobes, large
concave areas, and additional valleys, cliffs, and plains.
No assumptions for the active areas on the surface of \shortcomet{}
enter our model.

The surface is shown from different viewing directions in
Fig.~\ref{fig:water3d} and colored by the surface emission rate $\dot
\rho_{\HiiO,i}$ temporally averaged over three intervals,
respectively.
The first interval $A=(-330,-280)$ ends months before perihelion, the
second interval $B=(-50,50)$ covers the time around perihelion, and
the last interval $C=(340,390)$ begins months after perihelion.
According to Fig.~\ref{fig:qtimelinear} the dominating hemisphere
for the $\HiiO$ emissions changes from north in interval $A$ to
south in interval $B$ and back to north in interval $C$.

The integrated $\HiiO$ production over the complete interval
$(\misbegin{},\misend{})$ amounts to $780\pm 250$~kg/m$^2$ in the most
active source regions and to $110\pm 30$~kg/m$^2$ on average.
Assuming a pure water ice surface with a density of $470$~kg/m$^3$,
this corresponds to a maximum ice erosion of $1.7$~m.
The average ice erosion across the entire nucleus and orbit is then
$0.23$~m.
With increasing dust-to-gas ratio the erosion height increases
correspondingly.

To focus the discussion to regions of highest activity,
Fig.~\ref{fig:focusareas} shows the most abundant volatile $\HiiO$
around perihelion in the latitude/longitude Cheops-frame defined by
\cite{Preusker2015}.
Only those surface elements are depicted that contribute 50\% of the
total water loss during the time interval $B$.
Based on this set \nine{} oval activity areas are marked.
Area 1 covers parts of the regions Apis and Khonsu, area 3 parts of
the region Anuket, area 6 parts of the region Bastet, area 7 parts of
the region Bes and Khepry, area 8 parts of the region Bes and area 9
parts of the region Ash (see Fig.~11 of \cite{El-Maarry2016} for the
definition of regions).
Our activity areas contain 23 out of 34 locations of short living
outbursts around perihelion (small circles) reported by
\cite{Vincent2016}.
This remarkable correlation is even more pronounced and longer lasting
(including months before and after perihelion) in the $\COii$ data
discussed below.

The attached side panels to Figs.~\ref{fig:h2oco2} and \ref{fig:o2co} show the longitudinally averaged emission
(zonal mean) and in addition indicate the range of sub-solar latitudes during the considered interval.
Around perihelion and southern solstice (in interval $B$), all 
emission peaks are concentrated on the southern hemisphere close to
the sub-solar latitude at that time.
Months before inbound equinox (in interval $A$), the peaks for $\HiiO$
and $\Oii$ are also linked to the sub-solar latitude in the north.
Months after outbound equinox (in interval $C$), $\HiiO$ and $\Oii$
feature peaks near the northern sub-solar latitude but still have
contributions from the southern hemisphere.
In contrast to $\HiiO$ and $\Oii$, the peaks for the volatiles $\COii$
and $\CO$ are decoupled from the sub-solar latitude in the intervals
$A$ and $C$.
Substantial emissions originate from the southern hemisphere.
The strongest $\COii$ sources remain localized on the southern
hemisphere for all intervals independent to the corresponding
sub-solar latitude.

Figs.~\ref{fig:h2oco2} and \ref{fig:o2co} show the overall surface
emissions averaged within the time intervals $A$, $B$, and $C$ for all
species $\HiiO$, $\COii$, $\CO$, and $\Oii$.
For $\HiiO$ this corresponds to the three-dimensional representation
in Fig.~\ref{fig:water3d}.
The seasonally changing solar illumination leads to latitudinal
shifts in the source distribution, but with different patterns
for $\HiiO$, $\COii$, $\CO$, and $\Oii$.
Peak sources for $\HiiO$, $\COii$, and $\CO$ appear roughly at places
in agreement to \cite{Hoang2017} who projected the RTOF density
measurements to a $10$~km surface.
This agreement becomes even better when comparing the RTOF data for $\HiiO$
with Fig.~4 in \cite{Kramer2017}, which shows our inverse model data on a
$100$~km surface.
As suggested by VIRTIS-H observations in \cite{Bockelee-Morvan2016},
by modeling results in \cite{Fougere2016} and \cite{Hoang2017},
$\COii$, $\CO$ are decoupled from $\HiiO$ at the time before inbound
equinox.
% co2 /= h2o
This matches our observation in interval $A$, that $\COii$ and $\CO$ are
mainly located in the southern hemisphere, while $\HiiO$ originates from the
northern hemisphere.

Around perihelion (in interval $B$) the $\HiiO$ emissions are not
limited to the \nine{} activity areas but occur to some extent around
the entire nucleus.
$\CO$ and $\Oii$ are predominantly active in all water areas,
but $\COii$ coincides with water only for the southern areas 1-2, 4-8.
On the northern hemisphere, the $\COii$ emission is almost absent from
area 3, close to the Anuket fracture described in
\cite{El-Maarry2015a}, and area 9 in the Ash region.
Area 7 covers the patches reported by \cite{Filacchione2016} and
\cite{Fornasier2016} including high $\COii$ ice and $\HiiO$
ice concentrations around day $-145$ and around day $-105$, respectively.
Although their observations are made before our interval $B$, the
agreement for this source localization is still remarkable.

During the inbound northern summer (in interval $A$) $\HiiO$ and
$\Oii$ activity is located along a northern belt including the areas
3, 6, and 9.
This repeats in the outbound northern summer (in interval $C$) and is
complemented by activity in southern areas 1, 4, and 6 for $\HiiO$ and
in 1-2, 4-5, 7-8 for $\Oii$.
Thus, $\Oii$ source locations correlate to $\HiiO$ source locations
during all intervals $A$, $B$, and $C$.
For the inbound northern summer (in interval $A$) $\COii$ and $\CO$
activity is widely spread over the whole surface, $\COii$ exhibits
important contributions from the southern areas 1-2, 4-8, almost all
activity areas (except area 8) show $\CO$ emissions.
Comparing this pattern to $\HiiO$ sources, $\CO$ sources seem to
correlate to a linear combination of $\HiiO$ and $\COii$ sources.
At the same time despite the low emission from area 8, $\COii$
emissions in area 8 and surroundings in region Imhotep are still
higher than the $\HiiO$ emissions.
This shows a good agreement to the area of high ratio
$\rho_{\COii}/\rho_{\HiiO}$ described in \cite{Hassig2015a}.
During outbound northern summer, when $\mission{Q}{\HiiO}{}$ is almost
vanished, the pattern of $\CO$ sources seem to correlate to $\COii$
sources only.
Both source patterns focus to the southern areas 1-2, 4-8.

The $\COii$ sources are pinned to the south throughout the whole
Rosetta mission at the marked active areas:
for all intervals $A$, $B$, and $C$ the southern $\COii$ sources
(areas 1-2, 4-8) remain active.
This shows the consistent retrieval and assignment of $\COii$ sources
for the intervals $A$ and $C$, long before and after perihelion,
respectively.
Because these surface locations are reconstructed from completely
disjunct data sets and widely varying \scraft{} trajectories, this
validates our inverse model approach.
Furthermore, the location of $\COii$ sources on the southern
hemisphere is in agreement with the COPS data analysis for the month
May 2016 performed in \cite{Kramer2017}.

\section{Discussion}\label{sec:summary}

In this manuscript, we have presented emission rates for the gas
species $\HiiO$, $\COii$, $\CO$, and $\Oii$ with high spatial resolution on the
surface of \shortcomet{} and also temporally resolved in the time between
\datebegin{} and \dateend{}.
Previous surface maps were derived from lower resolution expansions
with 25 parameters by \cite{Fougere2016} and did not localize gas
sources due to the inherent averaging over longitudes.
The coma model by \cite{Marschall2017} considers various topographical
features as gas sources, does not employ an inversion process, and
leads to a non-unique source attribution.
The lower longitudinal resolution of the inversion models by
\cite{Hansen2016} (Fig.~10) and \cite{Fougere2016} (Fig.~5)
results in striped activity patterns and concentric fringes around
the poles, respectively.
With the 100~fold increase of resolution shown here, we obtain a more
accurate determination of local gas emitters on the surface,
validated by matching with independent optical observations of
outbreaks and spectroscopy of icy patches.
Another internal consistency check of the model is the assignment
of identical gas sources across completely distinct time-periods with
vastly varying solar radiation and spacecraft orbits.
In contrast to previous inversions, which work with single data sets
covering a long interval (300~days by \cite{Fougere2016a}), the
combined COPS/DFMS data set allows us to trace the coma evolution in
14~days intervals.
We also introduced a systematic uncertainty quantification due to
missing visibility of surface areas.
The reconstruction was based on the inverse gas model in
\cite{Kramer2017} and in situ DFMS/COPS measurements in the coma.
Based on the speed assumption in \cite{Hansen2016} for each of the
species, peak production rates (integrated over space) and integrated
(over space and time) productions rates are evaluated.
The summation over all gas species yields a peak production rate
$2.2\pm 0.1\times 10^{28}$~molecules/s, an integrated production rate
$5.8\pm1.8\times 10^9$~kg, and a maximum (averaged) water ice erosion of
$1.7$~m ($0.23$~m).
Incorporating the total mass loss, for the dust-to-gas ratio this
yields $0.5^{+1.1}_{-0.5}$.

\Nine{} activity areas are defined by $\HiiO$ emissions around
perihelion and those correlate well with short living outbursts
reported by \cite{Vincent2016}.
The examination of the \nine{} areas before, around, and after perihelion
shows that the source locations of $\HiiO$ and $\Oii$ follow the
sub-solar latitude and correlate to each other.
In contrast to that, $\COii$ sources are mainly located in southern
areas throughout the whole mission.
$\CO$ correlates to a linear combination of $\HiiO$ and $\COii$ months
before inbound equinox, months after outbound equinox it correlates to
$\COii$ only.

By comparing optical observations with dust-coma models
(\cite{Kramer2015,Kramer2018}) it is known that the dust coma is best
explained by a uniform activity across the entire sunlit nucleus,
which points to a rather homogeneous surface composition.

The surface localization of emissions for different gas species, also
described by \cite{AHearn2011} for comet Hartley 2, is a first step to
connect observational data to the reconstruction with first-principle
modeling of cometary activity such as suggested by
\cite{Keller2015a}.
The fast drop of the water production rates with increasing
heliocentric distance rules out the simplest sublimation models from
\cite{Keller2015a} taking a uniformly covered icy body with
$\mission{Q}{\HiiO}{} \sim \helio^{-2.8}$ in model A.
One way to accommodate higher exponents in the power law is to
consider a time-varying dust-cover on the surface, leading to a
transition from Keller model A to models with larger dust cover.
In addition, the peak water production of $\sim 3200$~kg/s in model A
(a completely water ice covered surface) is about five times as high
as our peak production.
A detailed comparison with first principle thermal and compositional
models of the surface is planned for future work.

\section*{Acknowledgements}

We thank H.U.~Keller and E.~K\"uhrt for helpful discussions.
The work was supported by the North-German Supercomputing Alliance
(HLRN).
Rosetta is an ESA mission with contributions from its member states
and NASA.
We acknowledge herewith the work of the whole ESA Rosetta team.
Work on ROSINA at the University of Bern was funded by the State of
Bern, the Swiss National Science Foundation, and by the European Space
Agency PRODEX program.

\label{lastpage}
\end{document}